\documentclass[fleqn,10pt]{wlscirep}
\usepackage[utf8]{inputenc}
\usepackage[T1]{fontenc}
\usepackage{lineno}
\usepackage{amsmath}
\usepackage{booktabs}
\usepackage{subcaption}
\usepackage{graphicx}

\usepackage[format=plain,justification=justified,singlelinecheck=false,font={stretch=1.125,small,sf},labelfont=bf,labelsep=space]{caption}

\newcommand{\ie}[0]{\textit{i.e.}, }

\newcommand{\eg}[0]{\textit{e.g.}, }

\usepackage{chemformula} 
\usepackage{braket}
\newcommand{\DefineAuthor}[2]{%
  \expandafter\newcommand\csname #1note\endcsname[1]{%
    \textbf{\textcolor{#2}{\textbf{#1:} ##1}}}%
  \expandafter\newcommand\csname #1\endcsname[1]{
    \textbf{\textcolor{#2}{##1}}}
  \expandafter\newcommand\csname #1cancel\endcsname[1]{%
    \textbf{\textcolor{#2}{\sout{##1}}}}%
  \expandafter\newcommand\csname #1change\endcsname[2]{%
    \textbf{\textcolor{#2}{\sout{##1} ##2}}}%
  \newenvironment{#1text}{\color{#2}}{\color{black}}
}
\definecolor{dartmouthgreen}{rgb}{0.05, 0.5, 0.06}
\DefineAuthor{LMS}{blue}
\DefineAuthor{LV}{red}
\DefineAuthor{AF}{brown}

\usepackage{xcolor}
\definecolor{AF}{RGB}{139,0,0}

\usepackage{dcolumn}
\usepackage{bm}

\title{On-the-Fly Machine Learning of Interatomic Potentials for Elastic Property Modeling in Al-Mg-Zr Solid Solutions}

\author[1]{Lukas Volkmer}
\author[1,*]{Leonardo Medrano Sandonas}
\author[2,3]{Philip Grimm}
\author[2,3]{Julia Kristin Hufenbach}
\author[1,4,*]{Gianaurelio Cuniberti}
\affil[1]{Institute for Materials Science and Max Bergmann Center of Biomaterials, TUD Dresden University of Technology, 01062, Dresden, Germany.}
\affil[2]{Institute of Materials Science, Technische Universität Bergakademie Freiberg, 09599 Freiberg, Germany}
\affil[3]{Leibniz Institute for Solid State and Materials Research Dresden, 01069 Dresden, Germany}
\affil[4]{Dresden Center for Computational Materials Science (DCMS), TUD Dresden University of Technology, 01062 Dresden, Germany}
\affil[*]{Corresponding authors: Leonardo Medrano Sandonas (leonardo.medrano@tu-dresden.de), Gianaurelio Cuniberti (gianaurelio.cuniberti@tu-dresden.de).}

\begin{abstract} 
The development of resilient and lightweight Aluminum alloys is central to advancing structural materials for energy-efficient engineering applications. 
To address this challenge, in this study, we explore the elastic properties of Al-Mg-Zr solid solutions by integrating advanced machine learning (ML) techniques with quantum-mechanical (QM) atomistic simulations. 
For this purpose, we develop accurate and transferable machine-learned interatomic potentials (MLIPs) using two complementary approaches: (i) an on-the-fly learning scheme combined with Bayesian linear regression during ab initio molecular dynamics simulations, and (ii) the equivariant neural network architecture MACE.
Both MLIPs facilitate the prediction of composition-dependent elastic properties while drastically reducing the computational cost compared to conventional QM methods. 
Comparison with ultrasonic measurements shows that the deviation between simulation and experiment remains within a few GPa across all Al-Mg-Zr systems investigated.
These potentials also enable the systematic exploration of the Al-Mg-Zr solid solution phase space and provide insights into the elastic behavior as a function of alloying element concentration.
Hence, our findings demonstrate the reliability and transferability of the parameterized on-the-fly MLIPs, making them valuable for accelerating the design of Al alloys with tailored physicomechanical properties in complex compositional spaces.
While the present study focuses on homogeneous phases, it establishes a foundation for future multiscale simulations that include microstructural features such as precipitates and grain boundaries.
\end{abstract}
\begin{document}

\flushbottom
\maketitle

\thispagestyle{empty}

\section{Introduction} \label{sec:intro} 
%
%
Aluminum (Al) alloys are widely used in industrial applications due to their low density, high strength, and excellent corrosion resistance. 
Their performance can be further enhanced through various strengthening mechanisms, including solid solution strengthening, work hardening, precipitation hardening, and grain refinement \cite{strengtheningexp1,strengtheningexp2}.
These properties make Al alloys especially valuable in the aerospace and automotive industries, where reducing weight is critical for improving energy efficiency and lowering fuel consumption \cite{application1,application2, application3}. 
High-strength 2xxx and 7xxx series alloys, with strengths exceeding 400 MPa, are particularly attractive for research and engineering in the field of additive manufacturing \cite{application4,application5, application6}. 
%
%
For example, Croteau \textit{et al.} demonstrated that Al-Mg-Zr alloys offer a remarkable combination of high yield strength and ductility \cite{AlloyingExp4}.
The addition of Zirconium (Zr) promotes the formation of the Al\textsubscript{3}Zr phase, which contributes to increased strength and is utilized in commercial additive manufacturing applications, such as Addalloy\textregistered\ \cite{AlloyingExp1}.
Compared to Scandium (Sc), which is often used for similar purposes, Zr provides a more cost-effective and sustainable alternative \cite{AlloyingExp2,AlloyingExp3}.
Magnesium (Mg) also plays a critical role by enhancing solid solution strengthening, thereby improving strength while maintaining ductility \cite{AlloyingExp1,AlloyingExp5}. 
However, incorporating two or more alloying elements into the Al matrix largely increases the number of potential compounds to evaluate in high-throughput studies. 
This complexity poses a challenge for experimental techniques in designing Al alloys with tailored physicomechanical properties.

To advance the exploration and understanding of mechanical properties in complex and previously uncharacterized Al alloys, analytical modeling and atomistic simulations have been widely employed in recent years.
%
%
These methods provide access to intrinsic material parameters that govern mechanical responses under defined conditions.
%
%
For instance, Chen \textit{et al.} developed a model in which Vickers hardness is expressed as a function of the shear and bulk modulus \cite{Vicker}. 
It has also been shown that the hardness, tensile strength, and intrinsic fracture strength of metals are related to their polycrystalline elastic constants  \cite{Pugh, Pettifor1, Pettifor2}. 
On the other hand, these properties can be derived from Density Functional Theory (DFT) or Molecular Dynamics (MD) simulations by applying small deformations and analyzing the resulting stress–strain or energy–strain behavior \cite{Neugebauer}. 
A DFT study reported that the bulk modulus of Al–Li intermetallics decreases with increasing Li content \cite{alloyingAl1}.
Similarly, both experimental and simulation data for Al–Cu alloys show a decreasing trend in elastic properties with increasing Al content\cite{alloyingAl3}. 
In addition, a growing database of DFT-calculated elastic tensors for over $1,000$ inorganic compounds has recently been compiled, enabling data-driven investigations of their elastic properties \cite{Charting}.
%
%
%
%
%

The temperature-dependence of physicomechanical properties can also be captured through MD simulations\cite{Sangiovanni1,Sangiovanni2}. However, these simulations rely heavily on the availability of accurate and transferable interatomic potentials to describe atomic interactions in alloys. 
Classical interatomic potentials are often system-specific, limiting their applicability for a broader exploration across compositional and configurational phase spaces.
This limitation can be addressed using ab initio MD (AIMD), in which atomic trajectories are derived from quantum mechanical calculations via DFT \cite{AIMD}.
Nevertheless, AIMD simulations become computationally expensive when applied to complex alloy systems with varying concentrations and configurations, where the use of large supercells containing hundreds of atoms are required.
%
%
To overcome these challenges, machine-learned interatomic potentials (MLIPs) have emerged as a promising alternative \cite{Schmidt19,Unke,Behler21,poma25}. 
These potentials harness the predictive capabilities of ML algorithms to model complex atomic interactions with near-DFT accuracy while drastically reducing computational cost.
Trained on DFT-generated datasets, these models can effectively capture key interatomic interactions across a broad range of compositions and configurations, enabling efficient exploration of chemical and mechanical property spaces. 
As a result, MLIPs have been successfully used to calculate mechanical properties in a wide variety of alloy systems \cite{mlffalloy1,mlffalloy2, mlffalloy3}. 
Additionally, large-scale materials property datasets have recently been developed (\eg Materials Project\cite{DatabaseMP}, Aflowlib\cite{DatabaseAFLOW}, Alexandria\cite{alexan}, NOMAD\cite{nomad}) and used to train universal MLIPs \cite{m3gnet,macemp0,omat24,Marques25}. 
Despite these advancements, a significant gap remains in the availability of structural and property data required to parameterize accurate and transferable MLIPs for ternary and quaternary Al alloys.

\begin{figure}[t]
\centering
\includegraphics[width=\linewidth]{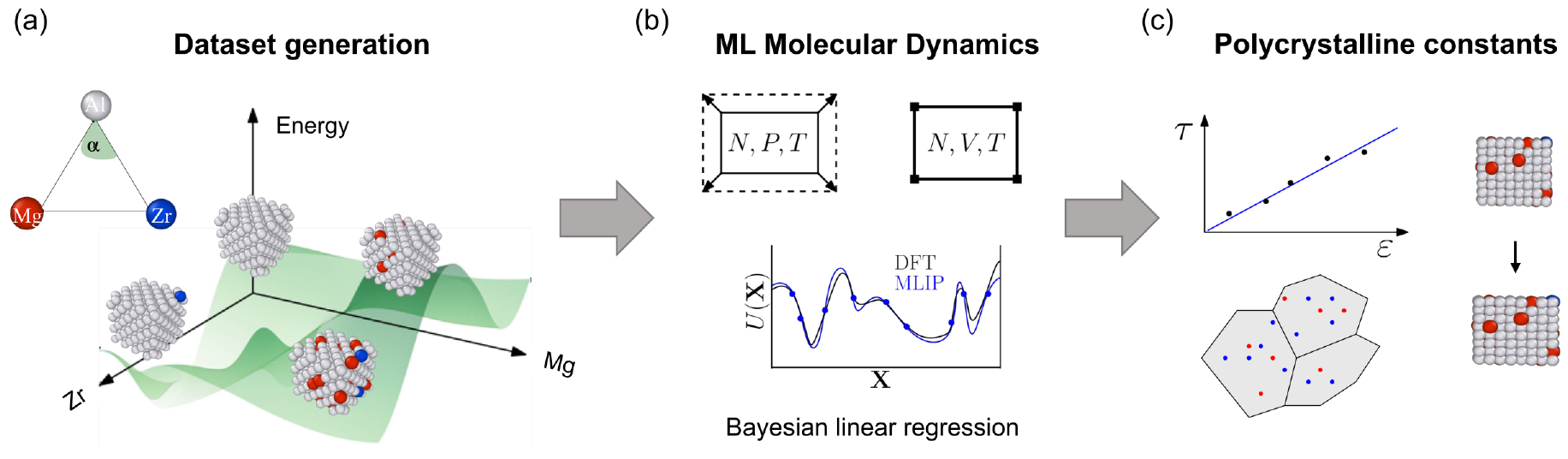} 
\caption{\label{fig:traininganalysis} Schematic representation of the data-driven strategy used for the fast and accurate derivation of elastic properties of Al-Mg-Zr alloys. (a) The solid solution phase space is sampled through structural optimizations at zero temperature using density functional theory (DFT). (b) The most favorable structures are then used to train a machine-learned interatomic potential (MLIP), employing an on-the-fly learning approach combined with Bayesian linear regression. (c) Finally, the elastic properties are obtained from stress-strain curves generated via ML-assisted molecular dynamics (MD) simulations.}
\end{figure}

Building on the context outlined above, this work investigates the elastic properties of oversaturated Al-Mg-Zr solid solutions by combining accurate atomistic simulations with ML techniques.
%
%
Specifically, we employ the on-the-fly learning method implemented in the Vienna Ab initio Simulation Package (VASP)\cite{VASP1,VASP2}, which utilizes Bayesian statistics and kernel ridge regression to efficiently parameterize reliable MLIPs with a reduced number of costly DFT calculations \cite{Jinnouchi1,Jinnouchi2,Jinnouchidescriptor}. 
Rather than focusing on temperature effects in multicomponent alloys\cite{Liu23,Sivaraman20,Bock24,Farache22}, we leverage this approach to develop a transferable MLIP capable of accurately predicting elastic properties across a wide range of alloy compositions and atomic configurations, including both binary and ternary Al systems (see Fig. \ref{fig:traininganalysis}). 
%
%
%
%
The dataset generated during the on-the-fly learning process is further used to train an additional MLIP based on the equivariant neural network architecture MACE \cite{MACE1,MACE2}, enabling a performance comparison between two state-of-the-art ML approaches.
Finally, from the stress–strain curves computed using both MLIPs (\ie VASP-ML and MACE-ML), we extract the stiffness matrix and subsequently derive the polycrystalline elastic constants.
%
%
These simulated values are then compared to experimental data obtained via ultrasonic measurements.
%
%

\section{Methods} \label{sec:met}

\subsection{Dataset generation}

Aluminum (Al) solid solutions form face-centered-cubic (fcc) structures. In supersaturated solid solutions, the equilibrium solubility of alloying elements is exceeded \cite{solidsolution}. 
This leads to greater solid solution strengthening, since the increase in yield strength is usually found to scale as $c^m$, where $c$ is the concentration of solutes and $m > 0$ is a parameter fitted to existing experimental data \cite{SSS}.
In this study, experimental values of solubility limits of Magnesium (Mg) and Zirconium (Zr) are used, which read 15$~$wt\% for Mg and 3~wt\% for Zr in oversaturated solutions \cite{Mgsolubility,Zrsolubility}. 
Zr has lower solubility due to its increased atomic radius and different bonding characteristics \cite{CahnHaasen}. 
Existing material databases usually do not contain structures for the solid solution space, \ie random atomic mixing in the unit(super) cell, and a vast number of possible configurations \cite{DatabaseMP,DatabaseAFLOW}. 
Indeed, these databases mainly focus on ordered, well-defined structures, which are easier to compute their properties and store.

Here, 3~wt\% of \ch{Zr} is equivalent to approximately 0.91~at\% in the \ch{Al} matrix. Accordingly, the atomic supercells investigated here contain 256 atoms, \ie $4 \times 4 \times 4$ unit cells of Al in order to represent the low amount of Zr.
To train the MLIP within the on-the-fly learning workflow implemented in VASP code \cite{VASP1,VASP2}, we begin with pure Al and progressively add other elements up to their respective solubility limits.
For a given concentration of Al, Mg, and Zr, $100$ different configurations are sampled, \ie the atoms are rearranged randomly. 
%
Each of the configurations is optimized at zero temperature, and afterwards the configurations with high and low energies are chosen for the next step, see Fig.\ S1 in the Supporting Information (SI). 
In a system with $N$ Mg atoms and $M$ Zr atoms, the total number of configurations scales as $\sim \frac{1}{N!M!(256-N-M)!} $ without symmetry considerations. 
Thus, the total energy spans over a wide range with increasing amounts of Mg up to the solubility limit.
%
%
Since the desired elastic properties are derived from a stress-strain analysis, deformed systems are also included in the training for the isothermal-isochoric ensemble (NVT-ensemble).
In doing so, the aforementioned training structures are distorted between values of $\delta = (-1\%, -0.5\%, 0\%,  +0.5\%, +1\%)$ as described in Sec. \ref{sec:elastic properties}.

All electronic structure calculations were performed using the VASP code \cite{VASP1,VASP2}. The projector augmented wave (PAW) method was employed to describe the electron-ion interactions \cite{PAW}. The Perdew-Burke-Ernzerhof (PBE) functional within the generalized gradient approximation (GGA) was used for exchange-correlation energy \cite{PBE}. The plane-wave basis set was truncated at a kinetic energy cutoff of 500~eV and the Brillioun Zone was sampled using a $\Gamma$-centered k-point grid with a spacing of 0.15~\AA\textsuperscript{-1}. For the electronic occupation functions Methfessel-Paxton smearing method with a smearing width of 0.3~eV was applied in all calculations. During structural relaxation at zero temperature, a conjugate gradient algorithm was employed with a convergence criterion of $10^{-6}$~eV for the total energy and 0.01~eV/\AA~ for the force on each atom. 

\subsection{Interatomic potential parameterization}
%
\subsubsection{Kernel-based active learning method}
As noted previously, machine-learned interatomic potentials (MLIPs) were developed using the on-the-fly learning method during ab initio molecular dynamics (AIMD) simulations, as implemented by Jinnouchi et al. in the VASP code \cite{Jinnouchi1,Jinnouchi2,Jinnouchidescriptor}.
%
To do this, a time step of 3~fs  was chosen,  optimized by finding out the Nyquist frequency of the dynamics \cite{Nyquist}. The isothermal ensembles were realized by employing a Langevin thermostat with a Langevin friction coefficient of 10~ps\textsuperscript{-1} for all atoms and lattice degrees of freedom \cite{Langevin1,Langevin2}. The temperature was set to 350~K in the training, and for application, the temperature was set to 298~K, which represents a subspace in the sampled potential energy surface and avoids thermal extrapolation in production runs \cite{Unke}.

%
%
Here, we briefly discuss the most important steps for training the MLIP, but additional information can be found in the SI.
The main assumption of this method is to approximate the total potential energy $U$ of $N_A$ atoms  in the system as a sum over local atomic energies $U_i$
\begin{eqnarray}
  U = \sum_{i=1}^{N_A}U_i = \sum_{i=1}^{N_A}\sum_{i_B=1}^{N_B} w_{i_B} K(\mathbf{x}_i, \mathbf{x}_{i_B}),
\label{eq:one}
\end{eqnarray}
where the latter are given as a linear combination of kernels $K$ with coefficients $w_{i_B}$. The descriptors are described in the SI and their values are represented by the vectors $\mathbf{x}$. In total,  $N_B$ reference atoms are chosen on the fly as described below.
%
In this model, a polynomial kernel of the form
\begin{eqnarray}
    K(\mathbf{x}_i,\mathbf{x}_j ) = ( \mathbf{\hat{x}}_i \cdot \mathbf{\hat{x}}_j)^\zeta
\end{eqnarray}
is used. The exponent $\zeta$ introduces mixing of two- and three-body descriptors, and it has been shown that mixing leads to a more accurate MLIP, especially for liquid systems \cite{Jinnouchidescriptor}. 
During the training, atomic forces and the stress tensor are calculated from derivatives of the potential energy. Hence, the potential energy, atomic forces, and stress tensor for the $\alpha$-th time step of an AIMD simulation are aligned to a $1+ 3 N_A + 6$ dimensional vector $\mathbf{y}^{\alpha} = (E^{\alpha}, \{\mathbf{F}_i^\alpha\}, \boldsymbol{\tau}^\alpha )  $ and Eq. \ref{eq:one} is written as a matrix equation, $\mathbf{y}^\alpha = \boldsymbol{\phi}^\alpha \cdot \mathbf{w}$,
where $\boldsymbol{\phi}^\alpha$ is a $(1+3N_A+6) \times N_B$ matrix containing terms like $ \sum_i K(\mathbf{x}_i^\alpha, \mathbf{x}_{i_B})$ and the corresponding derivatives. By introducing the super vector $\mathbf{Y} = (\mathbf{y}^1, \dots, \mathbf{y}^{N_{\mathrm{str}}})$ the previous linear problem is generalized to 
\begin{eqnarray}
    \mathbf{Y} = \boldsymbol{\Phi} \cdot \mathbf{w},
\end{eqnarray}
where $\boldsymbol{\Phi}$ is called design matrix. The $N_B$ linear coefficients are aligned to the vector $\mathbf{w}$. The desired optimal coefficients are calculated by Bayesian linear regression from all $N_{\mathrm{str}}$ structures  $\{\mathbf{y}^\alpha\} $ by maximizing the posterior conditional probability distribution\cite{Jinnouchi1,Jinnouchi2,Bishop}. 
%
%
The ML method is applied during an AIMD simulation. At each time step, the potential energy, atomic forces, and stress tensor are predicted by the trained MLIP.
Moreover, the error in the prediction of the forces is calculated using Bayesian statistics \cite{Bishop}.
This error is given by the uncertainty in the prediction of the new structure $\mathbf{y}$ based on already collected data $\mathbf{Y}$ as $p(\mathbf{y}|\mathbf{Y})$. 
If the maximum force error of any atom is above a certain threshold, the results for this time step are calculated by the selected DFT method. 
Otherwise, the MLIP is replacing the DFT calculation. The threshold for the Bayesian error is updated during the run based on the maximum force error for the last ten time steps. The MLIP is retrained by adding new local reference configurations when the Bayesian error is above the threshold, as described before. 
To reduce computational cost and avoid oversampling, the MLIP is not retrained after every DFT step. Furthermore, a CUR-algorithm is used to decide which local configurations are taken as reference \cite{Jinnouchi2,CUR}. 
The constructed MLIP of one AIMD simulation can be used for subsequent training in another AIMD simulation with a different concentration. 
This procedure is repeated until the MLIP is able to accurately describe the configurations of interest. 

\subsubsection{Equivariant neural network potential}\label{sec:mace}

In addition to training the MLIP using the kernel-based method in VASP, we also evaluate the performance of the MACE method \cite{MACE1,MACE2} on the Al alloys dataset generated on-the-fly in VASP. 
MACE is an E(3)-equivariant message-passing neural network (ENN) that generalizes the atomic cluster expansion, offering significantly improved efficiency and accuracy.
While ENN–based methods typically require more training data than kernel-based approaches, they often exhibit superior scalability and transferability when trained on complex datasets.
%
%
 Unlike traditional two-body message-passing schemes, MACE employs higher-order (up to four-body) messages, which allows capturing complex atomic interactions with just two message-passing iterations, substantially reducing training and inference time.
This represents a considerable advantage over typical graph neural networks requiring five or more iterations.
MACE constructs rotationally equivariant message functions via symmetrized tensor products of atom-centered neighbor representations. The readout network then predicts per-atom contributions to energy, forces, and stress, trained end-to-end by minimizing a similar combined loss function over energies, forces, and virials, just as in the kernel regression model. 
For detailed information about the architecture, the reader is referred to the original paper \cite{MACE1}. 
 The optimized hyperparameters of the MLIPs (VASP and MACE) are given in the SI. 


\subsection{Elastic properties} 

\subsubsection{Stress-strain relation}\label{sec:elastic properties}

In Hookean elasticity, the stress response $\tau_{ij}$ of a continuous material to applied strain $\varepsilon_{kl}$ is governed by a linear constitutive relation between the strain and stress tensor \cite{Poirier}. In contracted notation, this is given by
\begin{eqnarray}
    \tau_i = \sum_{j=1}^6 C_{ij} \varepsilon_j,
\end{eqnarray}
where $C_{ij}$ is the $6 \times 6$ stiffness matrix. Due to the symmetry of stress and strain tensor, the stiffness matrix contains 21 independent entries in the most general case.
For cubic structures, all components are zero except $C_{11}, C_{12}$ and $C_{44}$ \cite{Poirier}. In this work, the structures are stretched by employing lattice transformations $h_i' = \sum_j (\delta_{ij}+ \varepsilon_{ij}) h_j$, where $h_j$ and $\varepsilon_{ij}$ are undeformed lattice vectors and strain matrix, respectively. Due to the large system size and alloying, deviations from the minor and major symmetry of the stiffness matrix are expected. 
Therefore, the structures are stretched by employing three different lattice transformations given by $\varepsilon_i = \varepsilon_{i+3} = \delta  \, \, \,(i=1,2,3)$, where the other entries stay zero. From now on, we will use the Voigt notation \cite{voigt}. The stiffness constants are averaged as follows:
\begin{eqnarray}
    C_{11} &&= \frac{c_{11}+ c_{22}+c_{33}}{3}, \nonumber \\
    C_{12} &&= \frac{c_{12}+c_{13}+c_{21} + c_{23} + c_{31}+c_{32}}{6}, \\
    C_{44} &&= \frac{c_{44} + c_{55} +c_{66}}{3} \nonumber,
\end{eqnarray}
where $c_{ij}$ labels the actual entries in the stiffness matrix before symmetry considerations. Those can be derived by the ordinary least squares (OLS) method from the linear relationship between strain and stress. 
To improve the accuracy of our results, the following procedure is applied \cite{Charting}: firstly, two different sets of strain rates are deployed, $\delta_1 = (-1\%,-0.5\%,0\%, + 0.5\%, +1\%)$ and $\delta_2 = (-0.5\%,0\%, +0.5\%)$. Then, elastic constants are fit to both strain rates, and if the results differ more than 10\%, additional deformations at $\delta = \pm 0.25\%$ and $\pm 0.75\%$ are added. Otherwise, the elastic constants from $\delta_1$ are used to calculate the polycrystalline elastic constants. 

\subsubsection{Polycrystalline constants}\label{sec:poly}
Polycrystalline materials, composed of numerous randomly oriented grains, exhibit elastic properties that differ from those of their single grains. 
To characterize these properties, polycrystalline elastic constants are used, which represent averaged mechanical behavior over the bulk material. 
Understanding these constants is essential for predicting the performance of polycrystalline materials in engineering applications, where factors like grain size, texture, and boundary effects play critical roles \cite{VRH}
Two widely employed models for determining polycrystalline elastic constants are the Voigt model, which assumes uniform strain across grains, and the Reuss model, which assumes uniform stress. For cubic structures, these models yield \cite{Reuss,VRH},  
\begin{align}
    &B_V = B_R = \frac{C_{11}+ 2 C_{12}}{3}, \label{eq:bulk_modulus} \\
    &G_V = \frac{C_{11}-C_{12}+3C_{44}}{5},\quad 
    G_R = \frac{5(C_{11}-C_{12})C_{44}}{4 C_{44}+3 (C_{11}-C_{12})} \label{eq:shear_moduli}
\end{align}
 where $B$ and $G$ are the bulk modulus and shear modulus, respectively. The experimental moduli are expected to lie between the Voigt and Reuss values. Consequently, we use the Voigt-Reuss-Hill arithmetic average \cite{VRH2},
\begin{align}
    B_{VRH} = \frac{B_V + B_R}{2} , \quad 
    G_{VRH} = \frac{G_V + B_R}{2},
\end{align}
for a more accurate representation of the macroscopic elastic response of a material and, from now on, we suppress the indices. In isotropic materials, the elastic behavior is fully described by two constants, which in general can not be chosen uniquely \cite{VRH}. The Young's modulus $E$ and Poisson's ratio $\nu$ are given by the relations
\begin{align}\label{eq:enu}
    E = \frac{9BG}{G+3B},  \quad
    \nu = \frac{3B-2G}{2(3B+G)}.
\end{align}


\subsection{Experimental measurements}

\subsubsection{Sample preparation} \label{sec:sampleprep}

As a reference for ultrasonic measurements, pure Al (purity of 99.9995~\%, Haines \& Maassen, Germany) was used. 
%
%
%
The synthesis of alloys with different compositions (AlMg10, AlZr3, AlMg10Zr3 in wt\%) was utilized by centrifugal casting. Thereby, pure elements and pre-alloyed Al-Zr, produced by an arc melting device (Edmund Bühler, Germany), were placed into a glass carbon crucible and melted via an induction melting device (Linn High Term, Germany) under an argon atmosphere. Then, the melt is rapidly cast into a copper mold (12~mm diameter, 45~mm length). This technique provides high cooling rates (up to $10^{3}$~K/s) and rapid solidification of the material, which is essential for high saturation of the solid solution \cite{Philip1,Philip2,Philip3}.
 Finally, the cast rods were shaped into cubes with a length of approx.\ 10~mm by electronic discharge machining for further measurements. 

\subsubsection{Measurement of the moduli via ultrasonic waves}

The advantages of utilizing ultrasonic measurements over, \eg tensile tests, for determining the Young's modulus are a higher accuracy and a non-destructive methodology. As a device, the EPOCH 650 (Olympus, Germany) was used, which generates ultrasonic waves based on the reversed piezoelectrical effect.
Two different test heads (one for longitudinal and one for transversal waves) send ultrasonic waves through the material and record the time and intensity of ultrasonic echos. To couple the waves into the material, a glycerin-based and a highly viscous honey-like couple media for longitudinal and transversal waves were used, respectively. 
The different moduli can be determined by measuring the time between the first and second back wall echo of each ultrasonic wave (longitudinal and transversal). Several time measurements were carried out at room temperature with at least two specimens each and averaged. 
The velocity was calculated using the path-time law, based on the measured travel distance, which corresponds to the sample thickness.
Afterwards, the Poisson's ratio can be determined from the longitudinal $v_{long}$ and transversal wave velocity $v_{trans}$, which are related by,
        \begin{eqnarray}\label{eqexp1}
            v_{trans} = v_{long}\sqrt{\frac{1-2\nu}{2(1-\nu)}}.
        \end{eqnarray}
    Subsequently, the Young's modulus is calculated by,
        \begin{eqnarray}\label{eqexp2}
            E = v_{long}^2 \varrho \frac{(1+\nu)(1-2\nu)}{1-\nu},
        \end{eqnarray}
where $\varrho$ is the mass density of the material. The latter one was determined by the Archimedes' method using the density balance Sartorius MSA225s (accuracy: 0.01~mg) with a Cubis\textsuperscript{\textregistered } density kit (Sartorius AG, Germany), and the results are given in Table 1 of the SI. 
Finally, the bulk and shear modulus can be calculated from $\nu$ and $E$ by rearranging Eq.\ \ref{eq:enu}.

\subsubsection{Microstructure analysis}

The microstructure was investigated by scanning electron microscopy (Leo 1530 Gemini and Ultra Plus, both with SmartSEM software, Zeiss, Germany), in order to see other phases forming next to the $\alpha$-Al matrix and/or if segregation takes place. 
Energy-dispersive X-ray spectroscopy (EDX; Quantax400 with SDD-Detector Xflash4010, Bruker, USA) was performed to obtain elemental distribution maps and electron backscatter diffraction (EBSD; NORDLYS F detector, Oxford Instruments, United Kingdom) measurements to conduct phase maps. 
The applied electron voltage was 20~kV. The phases are indexed on the basis of the best-fit solution within the acquired EBSD images.
The following phases were indexed: cubic Al 
(ICSD-ID:~53772), cubic Al\textsubscript{3}Mg\textsubscript{2} (ICSD-ID:~57964) and tetragonal Al\textsubscript{3}Zr (ICSD-ID:~190892). Afterwards, outlier spikes and zero solution points were removed by the AZtec Crystal software (Oxford, UK) up to 5~\% cleaning.

\section{Results and Discussion} 

\begin{figure}[t]
\centering
\includegraphics[width=0.5\linewidth]{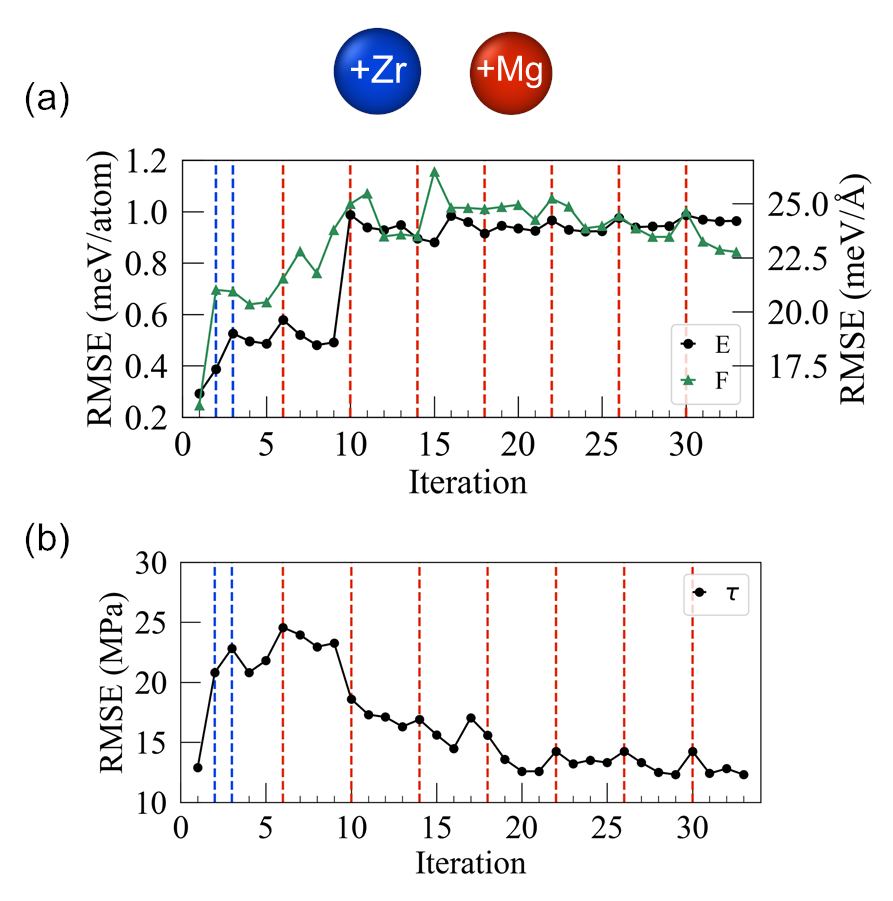} \\ 
\caption{\label{fig:2} Root Mean Squared Error (RMSE) over the entire training cycles with respect to the training data. (a) The RMSE values for total energies and forces are shown.
(b) The RMSE values for the stresses are displayed. Blue dashed lines indicate the point at which Zr is added to the training pool; red dashed lines mark the addition of Mg.}
\end{figure}

\subsection{On-the-fly learning scheme} \label{subsubsec:onthefly}

Two different VASP-ML models were parameterized using the on-the-fly learning method implemented in the VASP code.
The first model was trained in the isothermal-isobaric (NPT) ensemble and is used to relax structures to their pressure- and temperature-dependent volumes.
The second model, trained in the canonical (NVT) ensemble, is applied for stress-strain analyses. 
Details of the training structures for both models are provided in Table 2 in the SI. For the NVT-ensemble 2,693 reference structures were collected. Out of them 3,997 Al atoms, 3,179 Mg atoms and 1,286 Zr atoms are local reference atoms. The results for the NPT-ensemble are similar, but contained around 350 less reference atoms for Al and Mg. 
%
Fig. S2 in the SI shows the training history for a molecular dynamics (MD) simulation of Al\textsubscript{249}Mg\textsubscript{5}Zr\textsubscript{2}
During the initial 40\% of the simulation, the predicted Bayesian error frequently exceeds the adaptive threshold, indicating that the model encounters atomic environments not well represented by the current MLIP. 
Over time, as the model incorporates new local reference configurations, the predicted error decreases. Since the threshold itself evolves based on the history of the Bayesian error, it also gradually decreases.
A notable spike in the Bayesian error, exceeding 20~meV/\AA, occurs when new elements, specifically Zr and Mg, are introduced into the model. Consequently, the number of local reference configurations rapidly increases at the beginning of the training and eventually saturates.
A subsequent decrease in reference configurations results from sparsification by the CUR algorithm \cite{Jinnouchi2}. This training behavior was consistent across all cycles.
Throughout the entire training process, 98.2\% and 97.8\% of DFT calculations were bypassed for the NVT- and NPT-ensemble, respectively.
Fig.\ \ref{fig:2} provides a more detailed view of the evolution of the root mean square error (RMSE) for energies, forces, and stress tensor with respect to the training structures during the on-the-fly training. 
When Zr is added, an increase in RMSE is observed, reflecting the initial inability of the MLIP to describe interactions involving this new species. 
In contrast, the error increase due to Mg is less pronounced.
As new local environments are added to the training set, RMSE values for all properties decrease, demonstrating the adaptability of the VASP-ML model.
Notably, energy and force errors respond more quickly to the enriched dataset, while stress errors improve more gradually.
These results underscore that each major compositional change initially challenges the model but is followed by a convergence phase where prediction errors stabilize. 
This demonstrates the effectiveness of the on-the-fly learning method in accurately capturing interatomic interactions across diverse alloy compositions and atomic environments.
%
%
Final RMSE values for all properties are listed in Table \ref{tab:rmse}, confirming successful model adaptation and convergence.
%

To complement the method implemented in VASP, we also trained an alternative model using the equivariant neural network architecture MACE, as briefly described in Sec. \ref{sec:mace}.
The MACE-ML model was trained on the reduced training dataset that was generated during the on-the-fly learning procedure in NVT-ensemble.
%
%
This dataset considers $2,693$ non-equilibrium conformations of $34$ atomic configurations of Al-Zr-Mg systems. 
While the VASP-ML model serves as the primary model in our study, the use of MACE provides a valuable cross-validation benchmark and enables us to assess the generalizability of the training data generated by the first MLIP. 
This approach allowed us to isolate the influence of the underlying ML method on the quality of the resulting MLIP, without introducing variations due to changes in the dataset.
Details of the hyperparameter optimization of the MACE model can be found in Table 4 of the SI.
The best-performing MACE-ML model achieved RMSE values of 2.3 meV/atom, 16 meV/\AA{}, and 0.18 GPa for the prediction of energies, forces and stress in the training set.
The errors in energies and forces are comparable to those of the VASP-ML model; however, the stress prediction shows a larger deviation, which may impact the accuracy of predicted elastic properties.

\begin{table}[t]
    \centering
    \caption{Root Mean Square Error (RMSE) values for energies, forces, and stresses evaluated on the Al-Mg-Zr systems in the training set and test set. For the VASP-ML model trained in the NVT ensemble, three different weights were applied to the stress contribution in the loss function. Additionally, RMSE values for the best-performing MACE-ML model, evaluated on the same test set, are provided. Since the stress tensor consists of six independent components, only the RMSE corresponding to the largest component is reported. The units for energy, forces and stresses are meV/atom, eV/\AA \, and GPa, respectively. The bold numbers indicate the models, which were used for the calculation of elastic properties. } 
    \setlength{\tabcolsep}{2pt}
    \renewcommand{\arraystretch}{1.5}
    \begin{tabular}{ccccccccc}
    \toprule
&& &&Training data & &&Test data & \\
\midrule
\textrm{Model}&
&
\textrm{$\omega_\tau$} &
\textrm{E}&
\textrm{F}&
\textrm{$\tau$}&
\textrm{E}&
\textrm{F}&
\textrm{$\tau$}\\
\midrule
 VASP-ML& NVT &1&1.8& 0.022& 0.02  & 2.0 & 0.031 & 0.49 \\
 &  &\textbf{10}& \textbf{0.9}&\textbf{0.023}&\textbf{0.01} & \textbf{2.3 }& \textbf{0.044} & \textbf{0.02} \\
 &&100 &3.6&0.059&0.02& 39.3 &0.120 & 0.02\\
& NPT &10&1.1 &0.020&  0.01 & 4.0 & 0.047 &0.01 \\
  MACE-ML&  & &\textbf{2.3}&\textbf{0.016}&\textbf{0.18} &\textbf{2.7} & \textbf{0.026} & \textbf{0.19} \\
    \bottomrule
    \end{tabular}
    \label{tab:rmse}
\end{table}

\begin{figure}[t]
\centering 
\includegraphics[width=0.52\linewidth]{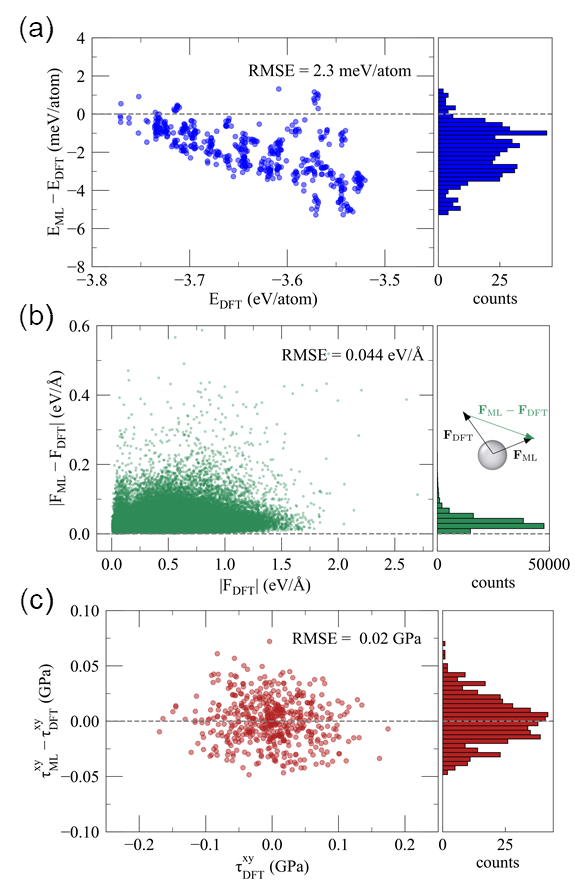}
\caption{\label{fig:validation} Comparison between DFT and ML results of (a) total energies, (b) forces, and (c) stresses for Al-Mg-Zr systems in the test set. 
%
For the forces, the absolute value of the vectors for each atom were calculated as illustrated. 
One entry of the stress tensor is presented, the RMSE for the full stress tensor is given Table 3 of the SI. Next to each graph, histograms illustrate how the properties are distributed around zero difference (indicated by grey dashed lines).
}
\end{figure}

\subsection{Transferability to unseen chemistries}

To ensure the reliability and accuracy of the developed MLIPs, different validation benchmarks were carried out. The goal is to assess the ability of these potentials to predict elastic properties of Al solid solutions with different amount of Mg and Zr up to the solubility limit. 
As a first step, we compare DFT data with predicted data (energies, forces and stresses), including stretched systems in the elastic regime. 
We generated 100 different structures containing Mg and Zr, excluding the atomic compositions used during the training procedure. 
This ensures that the phase space up to solubility limit is sampled properly, \ie we spread the concentrations over the whole solid solution phase space.

MD simulations at 298~K were performed using both trained MLIPs, \ie VASP-ML and MACE-ML.
Atomic configurations were sampled at equally spaced time steps, and single-point calculations using the VASP code were carried out to generate the test set for the MLIP study.
%
Fig. \ref{fig:validation} shows the residual plots for energy, forces, and stress of each snapshot, comparing results from the VASP-ML model with DFT calculations. 
%
%
The DFT energies and stresses vary due to differences in alloying content and strain within the test set. 
As energy increases, the deviation of ML predictions also increases slightly---an effect primarily caused by the presence of strained structures.
Accordingly, the model shows better accuracy for equilibrium structures ($\varepsilon = 0$). Even in strained cases, however, the maximum energy deviation was approximately 6 meV/atom.
Since elastic properties are derived from stress responses, particular attention is given to the accuracy of the stress tensor.  Typical engineering stress-strain curves for structural Al alloys (\eg 5xxx series) exhibit yield strengths on the order of 100–500 MPa, marking the limit of the elastic regime \cite{stressstrainAlalloy}. 
Thus, precision in the range of MPa is required for accurate derivation of elastic constants.
In ML models, the relative weight of different properties can be adjusted during training to prioritize specific quantities.
 This technique can enhance the accuracy of stress predictions but may reduce the accuracy of energies and forces \cite{Liu, Unke}.
To evaluate this trade-off, we trained three models for the NVT ensemble using stress weights $\omega_\tau$ set to 1, 10, and 100, while keeping the weights for energies and forces constant.
The RMSE values for the test set are reported in Table \ref{tab:rmse}.
A stress weight of $\omega_\tau = 10$ significantly improved stress tensor accuracy, with only a slight increase in energy and force errors.
In contrast, setting $\omega_\tau = 100$ resulted in a substantial increase in energy and force errors without further improving stress accuracy. 
Based on these findings, we used $\omega_\tau = 10$ for both the NVT and NPT ensembles, as the resulting RMSE values were sufficiently accurate for stress-strain analysis of our systems.

The performance of the MACE-ML model was evaluated on the same test set, and the RMSE values are listed in Table \ref{tab:rmse}. We found that the RMSE in both energies and forces were comparable to those obtained with the VASP-ML models, indicating that the training data carried sufficient information for accurate parameterization of interatomic potentials across different ML methodologies. However, the RMSE in stresses is one order larger in the MACE model. 
We want to remark, that including the stress tensor in the training data leads to a reduction of the RMSE for stress prediction in comparison to MACE models trained only on energies and forces.
In Table 4 of the SI, the RMSE for MACE models with different hyperparameters  (either with or without stress included in the training process) is given.

\begin{figure}[t!]
\centering 
\includegraphics[width=0.9\linewidth]{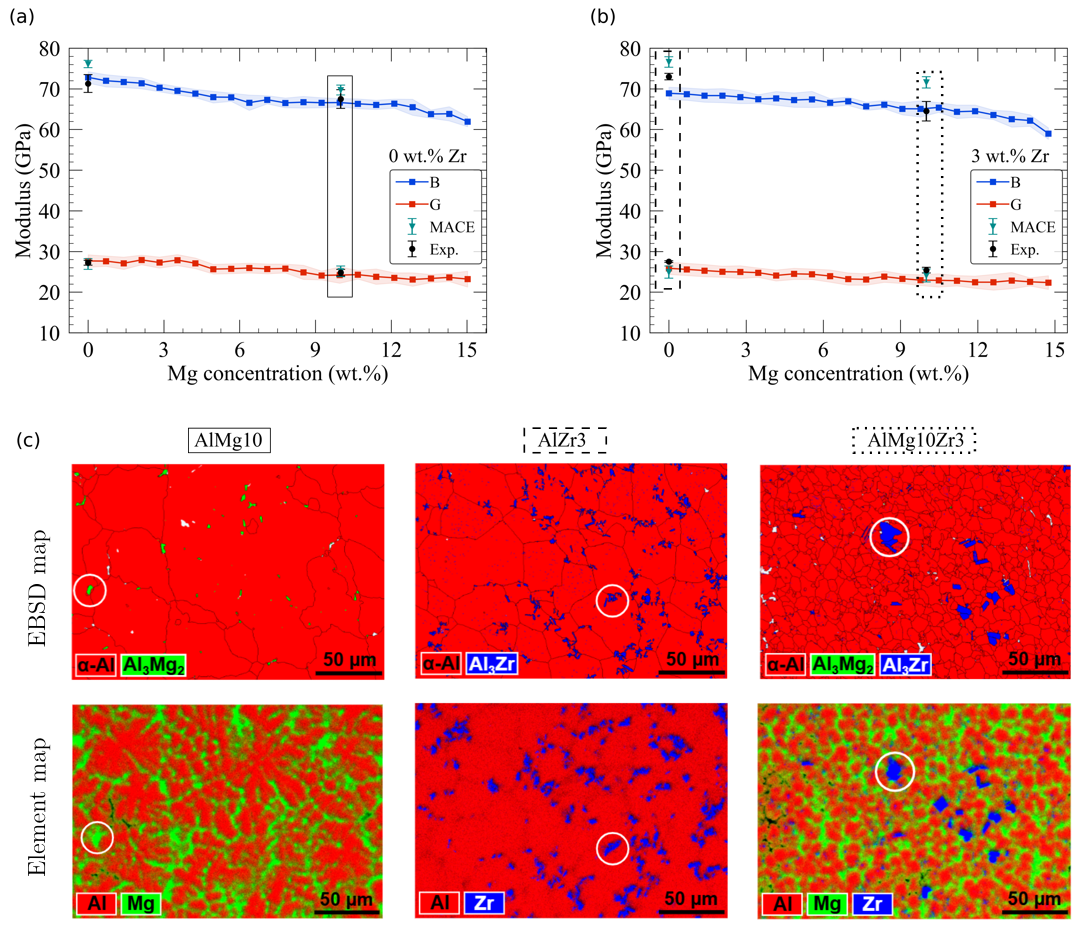}
\caption{\label{fig:moduli} Variation of the bulk modulus (blue line) and shear modulus (red line) as a function of Mg concentration for (a) alloys without Zr and (b) alloys with the maximum Zr content.
Results predicted by MACE-ML are indicated by green triangles. For comparison, the elastic constants of four experimentally synthesized Al alloys (\ie Al, AlMg10, AlZr3, AlMg10Zr3) determined via ultrasonic measurements, are shown as black circles. (c) Element and EBSD phase maps of AlMg10, AlZr3, and AlMg10Zr3 samples synthesized by centrifugal casting.
}
\end{figure}

\subsection{Predicting elastic properties of binary and ternary alloys}
To screen the entire ternary solid solution Al-Mg-Zr phase space, we generate structures up to the experimental solubility limits \cite{Mgsolubility, Zrsolubility}. 
Modeling random solid solutions is challenging, as the limited system sizes in methods like DFT often fail to capture true randomness due to periodic boundary conditions.
Thus, we applied the Special Quasirandom Structures (SQS) approximation for generating the configurations \cite{icet,sqs1,sqs2}.
Afterwards, to reflect the conditions of the experiments, the structures are equilibrated for 30~ps in the NPT-ensemble at a temperature of 298~K and a pressure of 1~Bar, \ie ambient conditions, ensuring volume, temperature and pressure convergence  \cite{Frenkel}. 
Afterwards we average the lattice constants over the last 500 steps of the simulation.
From this we model the strain by applying distortions given by three different strain matrices   and for each of those, multiple increments $\delta$ are used as described in Sec. \ref{sec:elastic properties}.
For each of the strained samples, we use the VASP-ML model for NVT-ensemble and run the dynamics for 30~ps, \ie we fix the volume to the strained structure. 
The alloys encounter stress due to the lattice changes.
The stresses $\tau_{i}$ are averaged over the last 500 steps and then the OLS method is used for the regression analysis of elastic constants. 
The simulation time is justified by analyzing the behavior of pressure, volume and temperature during the simulation, \ie those quantities should have negligible fluctuation in the respective ensemble \cite{Frenkel}.

We now calculate the bulk modulus, shear modulus, Young's modulus and Poisson's ratio using the VASP-ML, as it was described in Sec. \ref{sec:poly}. 
Fig.\ \ref{fig:moduli} presents the evolution of the bulk and shear modulus as a function of the Mg concentration for Al-Mg alloys containing either no or maximal amount of Zr. 
The largest bulk elastic constant was received for pure Al with 72.8~GPa, which is in good agreement with reference results from a DFT-study using generalized gradient approximation (GGA) \cite{B-Al-DFT}. 
The bulk modulus decreases with increasing amount of Mg down to 59.0~GPa for AlMg15Zr3. 
Small fluctuations come from numerical noise. The reduction in bulk modulus can be attributed to the atomic size mismatch and the resulting lattice strain induced by Mg atoms. 
%
The decreasing behavior is the same for the shear modulus although it is less affected than the bulk modulus, \ie the solid becomes more compressible but retains more of its resistance to shear deformation. 
Alloying with Zr affects the polycrystalline constants to a similar extent as alloying with the same concentration of Mg.

\begin{figure}[t]
\centering 
\includegraphics[width=\linewidth]{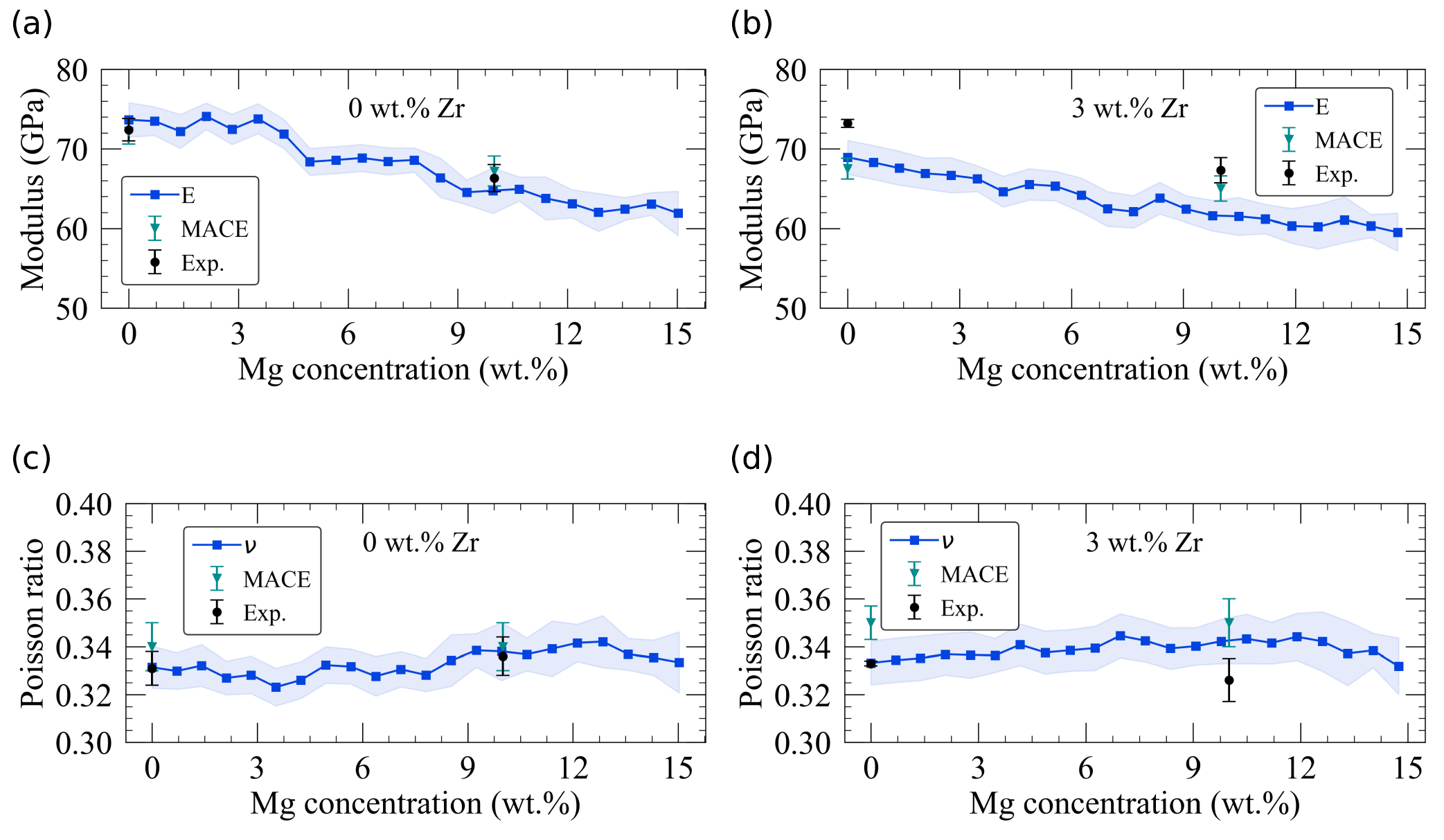}
\caption{\label{fig:2x2grid} Variation of the (a,b) Young's modulus and (c,d) Poisson's ratio as a function of Mg concentration for alloys without and with the maximum Zr content.
These properties were calculated from the bulk and shear modulus.
Results predicted by MACE-ML are indicated by green triangles. For comparison, the experimental values of Al, AlMg10, AlZr3, and AlMg10Zr3 are shown as black circles. 
}
\end{figure}

Fig. \ref{fig:2x2grid} shows the additional elastic properties calculated from the bulk and shear modulus. 
The decreasing trend with respect to the Mg and Zr amount is preserved for the Young's modulus, where it has its lowest value of 59.8~GPa for the AlMg15Zr3 system. 
However, there are some deviations from this behavior such that we have the largest Young's modulus of 66.9~GPa for AlMg2.25. 
The Young's modulus is sensitive to  uncertainties of the bulk and shear modulus. Therefore fluctuations of several GPa are present, especially for the Al-Mg system with low amount of Mg. 
We would like to emphasize, that the deduction of the Young's modulus from the other elastic properties (as described in  Sec. \ref{sec:poly}) implies two independent elastic constants. This kind of assumption holds, if the cubic crystallinity of pure Al is maintained after alloying. 
Even if alloying with such small amounts of these metals does not induce a phase transition, increasing the solute concentration can lead to distortion effects, especially considering that the equilibrium crystal structures of both Mg and Zr are hexagonal \cite{materialsproject}.
Therefore we analyzed the symmetry of the solid solutions utilizing the \texttt{spglib} package \cite{spglib}.
For the studied systems sampled by the SQS approach, no remarkable deviations from the cubic structure occurred, and consequently we abide by the description for cubic crystals.
The Poisson's ratio is fluctuating around 0.33 and in the typical range for conventional Al alloys \cite{Poissonexp1,Poissonexp2}. The variations are not following a trend with a maximal change of approximately 0.02. Compared to the elastic moduli, the Poisson's ratio is way less affected by alloying.
This suggests that the volumetric-to-shear deformation behavior remains relatively stable despite changes in composition. The arguments for the fluctuation behavior are the same like for the Young's modulus as described before. 
The uncertainties (represented as error bars in Fig. \ref{fig:moduli} and Fig. \ref{fig:2x2grid}) were calculated using error propagation, as described in the SI.

Finally, we computed the elastic properties of the four representative experimental compositions (see Sec. \ref{sec:rexp}) using the MACE-ML model. These results, along with values obtained using VASP-ML, are summarized in the Table S5 in the SI.
As shown in Fig.~\ref{fig:moduli}, the bulk moduli predicted by MACE-ML are slightly higher compared to those from the kernel-based MLIP, although the overall decreasing trend across compositions is preserved.
The shear and Young's modulus show similar behavior for both MLIP approaches. 
However, the Poisson ratio is overestimated, up to 0.35, primarily due to the increased bulk modulus.
The qualitative agreement between the two MLIPs indicates that both methods approximate the underlying potential energy surface consistently, despite their methodological differences. 

\subsection{Validation against experimental data} \label{sec:rexp}

To validate the computational study, a series of experimental investigations were carried out on pure Al and selected Al-based binary (AlMg10, AlZr3) and ternary alloy (AlMg10Zr3) systems. 
These alloys were produced under controlled conditions to ensure high quality and compositional accuracy, as described in Sec. \ref{sec:sampleprep}. 
Ultrasonic testing was employed to determine the elastic properties in a non-destructive manner, while additional microstructural characterization was performed using electron microscopy and diffraction-based techniques. 
This combined approach enabled both the macroscopic mechanical behavior and the underlying phase formation to be assessed and provided a solid experimental foundation for comparison with the theoretical predictions.

The material density and four polycrystalline elastic constants are given in Table S1 of the SI. The density of pure Al decreases from 2.696(1)~g/cm³ to 2.576(10)~g/cm³ when Mg is alloyed. 
If Zr is added to the material systems, the densities slightly increase by approximately 0.02-0.05~g/cm³. The standard deviations are larger for the alloys than for pure Al. 
This might be due to pores and voids appearing in casting during solidification shrinkage. Those defects would also lead to a lower measured density than the actual density.
The Poisson ratio is almost comparable for all investigated samples and lies around 0.33 (see black dots in Fig. \ref{fig:2x2grid}). The experimental results revealed that alloying with Mg tends to reduce stiffness (Young's Modulus) by approximately 6~GPa, while Zr contributes to a slight strengthening effect about 1~GPa.
Notably, the combination of both elements results in a balance of these effects. The errors are roughly 2~GPa for pure Al and when Mg is involved. 
AlZr3 shows a smaller error. The larger deviation for Al-Mg systems might occur due to evaporation of Mg during casting leading to a higher amount of trapped gas pores inside the sample. 
Those pores lead to a higher noise in the ultrasonic signals. Furthermore, pure Al shows also a larger error due to textured microstructure.

The EBSD and element maps of the microstructure of the alloyed systems are displayed in Fig. \ref{fig:moduli}c. 
Initially, the topography maps reveal a gray matrix, black regions corresponding to voids, and bright rod-shaped or cubic precipitates (see Fig. S3 in the SI).
To identify the phase composition, element maps were measured, where it shows that the gray matrix contains Al while the bright phase includes Zr. 
%
Additionally, Mg is distributed inhomogeneously, segregating along an interdendritic network. The Al dendrites form the primary substructure.
Despite the observed Mg segregation, the phase maps indicate that only a small amount of the intermetallic \ch{Al3Mg2} phase is present. This suggests that most of the Mg is dissolved within the Al matrix (see Fig. \ref{fig:moduli}c). 
Furthermore, the tetragonal \ch{Al3Zr} is present in both Zr-containing alloys and represents the bright phase. 
The elemental maps also show that Zr is primarily concentrated in these precipitates, implying that only a limited amount of Zr is dissolved in the Al matrix. Nevertheless, the matrix may still be saturated with Zr to a certain extent.
%
%
Overall, this microstructural analysis confirms the formation of a solid solution within the Al matrix, along with minor intermetallic phases that vary depending on the alloy composition. These findings will support and help explain the results obtained from the atomistic simulations.
%

%
%
By comparing the experimental results with those obtained from the VASP-ML and MACE-ML models, it is evident that the elastic properties of pure Al are accurately reproduced (see Fig. \ref{fig:moduli} and Fig. \ref{fig:2x2grid}).
AlMg10 also shows good agreement with experimental values, while that of AlZr3 and AlMg10Zr3 is slightly underestimated in VASP-ML. 
MACE-ML predicts slightly higher bulk modulus values compared to experiments, although it correctly captures the decreasing trend.
The absolute differences in shear modulus between experiments and MLIPs are smaller; however, since shear modulus is less affected by alloying, the relative deviation is comparable to that observed for the bulk modulus.
Overall, considering the error margins, the simulated and experimental values for both bulk and shear modulus are in good agreement.
As previously mentioned, the Young's modulus and Poisson's ratio were derived assuming isotropic behavior and calculated from the other elastic constants. Consequently, the uncertainties propagate, resulting in greater fluctuations in these values as a function of concentration.
The MLIPs used in this study were trained and applied strictly within the solid solution regime. Therefore, the computed elastic moduli represent the intrinsic properties of a homogeneous polycrystalline matrix and do not account for secondary phases, precipitates, or inclusions.
Referring to the phase maps in Fig. \ref{fig:moduli}c, secondary phases are indeed present in the experimental microstructures. 
In the Al-Mg system, a significant amount of the \ch{Al3Mg2} phase is observed, while in the Al-Zr system, \ch{Al3Zr} precipitates are present. In the ternary alloy, both phases coexist alongside the matrix.
When comparing our simulation results to literature data, the elastic moduli of \ch{Al3Zr} are reported to be significantly higher \cite{FPAl3Zr,ExpAl3Zr}, whereas those of \ch{Al3Mg2} are slightly lower \cite{Al3Mg2,Al61}.
%
%
%
%
%
Nonetheless, our simulated results using VASP-ML align more closely with the experimental values reported in this study, even though the measured properties are effective values influenced by the microstructure. 
Specifically, for AlMg10, only a small discrepancy is observed between simulation and experiment, while in the presence of \ch{Al3Zr} precipitates, the deviation remains within a few GPa.

%

\begin{figure}[t]
\centering 
\includegraphics[width=0.5\linewidth]{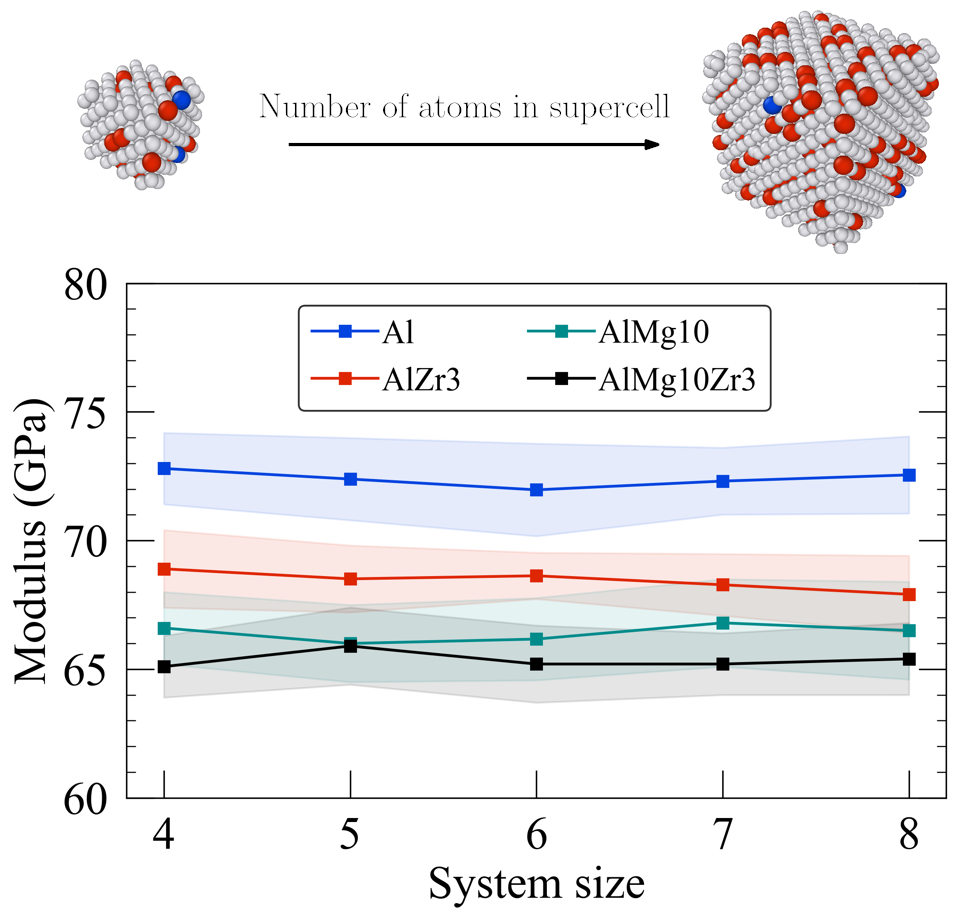}
\caption{\label{fig:scalability} Scalability benchmark for calculating the bulk modulus using the VASP-ML model. For this analysis, we considered four experimentally synthesized Al alloys. For each alloy, different system sizes were evaluated. The number on the x-axis indicates the number of unit cells along each direction of the constructed supercell.
%
}
\end{figure}

\subsection{Scalability benchmark}

The VASP-ML model was initially trained on configurations derived from $4 \times4 \times 4$ supercells (containing 256 atoms, ca.\ 16.3~\AA{} in length). 
All results discussed in the previous sections were evaluated using this system size. 
To assess the generalization capability of the VASP-ML model, we extended the evaluation to larger supercells beyond the original training scale.
Accordingly, the elastic modulus of systems with sizes up to eight unit cells in each dimension was computed.
This analysis was conducted for the four alloy compositions that were experimentally investigated (see Sec. \ref{sec:rexp}). 
The largest simulation cell therefore contained $2,048$ atoms and measured ca. 32.6 \AA ~ in each direction---dimensions that are generally inaccessible using pure DFT calculations.
The structures were generated using the SQS approach. To determine the elastic properties, we followed the same procedure as before: first equilibrating the structures in the NPT ensemble using the corresponding MLIP, followed by NVT simulations under varying stretch states.
The weight percentages of Mg and Zr were preserved across all system sizes, and the exact numbers of alloying atoms for each system size are provided in Table 6 of the SI. 
Throughout the MD simulations, the structures remained intact and exhibited no anomalous behavior.
In Fig.\ \ref{fig:scalability}, the bulk  modulus for the four compositions is shown as a function of system size. 
In each case, the bulk modulus fluctuates by around 1 GPa with respect to supercell size. This small variation indicates that the predicted modulus remains stable across different system sizes.
Taken together, these findings indicate that the VASP-ML model reliably captures the underlying interatomic interactions and is transferable to larger-scale systems.

\section{Conclusions} 

In conclusion, we have parameterized ML interatomic potentials (MLIPs) to accelerate accurate atomistic simulations of unexplored Al-Mg-Zr alloys under different mechanical conditions.
Because calculating elastic properties from ab initio MD data is computationally expensive for large supercells, we combined on-the-fly learning with quantum-mechanical methods to develop accurate and transferable MLIPs using reduced training sets, significantly improving efficiency.
Specifically, we parameterized the VASP-ML and MACE-ML potentials using Bayesian kernel ridge regression and the equivariant neural network architecture MACE, respectively.
Both MLIPs were employed to compute the polycrystalline elastic constants of supersaturated Al solid solutions with Mg and Zr concentrations up to the solubility limit, yielding consistent results. The MACE-ML potential exhibited a systematic but small overestimation. 
Our simulations also showed that the elastic constants decrease with increasing Mg and Zr content, which can be attributed to distortion effects within the Al matrix.
%
The scalability tests of the VASP-ML potential demonstrated robust performance, highlighting its suitability for large-scale simulations of realistic structures, including those with precipitates or local compositional heterogeneities.

%
%
%
%
%
%
%

Notably, the calculated elastic properties of the synthesized Al-Mg-Zr solid solutions are in good agreement with the experimental results, with the exception of the AlZr3 system. 
These deviations may arise from the presence of multiple structural phases in the microstructure of the alloy, as revealed by the measured EBSD maps.
As a potential way to mitigate this problem, a suitable homogenization technique that incorporates contributions from all phases (\ie the matrix and precipitates) can be applied, allowing our simulations to more accurately reproduce the experimental outcomes.
In brief, we demonstrated that by leveraging on-the-fly learning using DFT reference data, we were able to develop computationally efficient interatomic potentials to explore the Al-Mg-Zr solid solution phase space and compute accurate elastic properties. 
 %
%
Moreover, the MLIPs are expected to be extendable to other alloying elements, \eg substituting Zr with Si, which is known to improve yield strength in Al-Mg-Si alloys \cite{Silicon}.
To achieve full predictive capability for bulk mechanical behavior or component-level performance, future work should extend our MLIPs to include multiphase systems and interface effects, either by expanding the training dataset or through hierarchical modeling approaches.
The computed elastic constants can then serve as key input parameters for effective property calculations in multiscale modeling frameworks, including finite-element and phase-field methods \cite{KochmannMultiscale,FEM,Phasefield}. 
Optimizing these data-driven methods with experimental data will further enhance the design of advanced technological materials, such as mechanical metamaterials\cite{Jiao23}, by providing rapid and accurate input for geometry and topology refinement at both micro- and mesoscale levels.

\section*{Acknowledgments}
This work was supported by
the German Research Foundation (DFG) within the
Research Training Group GRK 2868 “Data-Driven
Design of Resilient Metamaterials” (D\textsuperscript{3}) – project
number 493401063. 
We thank the Center for Information Services and
High-Performance Computing (ZIH) at TU Dresden for providing the computational resources to obtain the results presented in this work. 


\section*{Data availability}
The dataset files containing the training and validation data of Al-Mg-Zr alloys are available in the  Github repository \href{https://github.com/lmedranos/MLalloyD3}{Repo-MLalloyD3}.
The models, codes, and additional benchmarking datasets used in this work can also be found in \href{https://github.com/lmedranos/MLalloyD3}{Repo-MLalloyD3}.


\section*{Author contributions}

The work was initially conceived by LV and LMS, and its design was developed with input from PG, JKH, and GC. LV and LMS trained the MLIPs using VASP and MACE, and conducted performance analyses. LV also evaluated the capabilities of MLIPs to compute the elastic properties of Al alloys. PG synthesized the Al alloys and performed all experimental measurements. LV and LMS drafted the original manuscript. All authors discussed the results and contributed to the final version of the manuscript.

\section*{Competing interests}

The authors declare no competing interests.

\bibliography{sample}


\end{document}


\fancyhead{}
 \renewcommand{\headrulewidth}{1pt}
 \renewcommand{\footrulewidth}{1pt}
 \setlength{\arrayrulewidth}{1pt}
 \setlength{\columnsep}{6.5mm}
 \renewcommand{\figurename}{Supplementary Figure~\!\!}
 \renewcommand{\tablename}{Supplementary Table~\!\!}

\renewcommand{\refname}{Supplementary references}

 \begin{center}
 \noindent\LARGE{Supplementary Information (SI) for:}
 \vspace{0.3cm}

 \noindent\LARGE{\textbf{On-the-Fly Machine Learning of Interatomic Potentials for Elastic Property Modeling in Al-Mg-Zr Solid Solutions}}
 \vspace{0.6cm}

 \noindent\large{\textbf{Lukas Volkmer,\textit{$^{1}$} Leonardo Medrano Sandonas,\textit{$^{1,*}$} Philip Grimm,\textit{$^{2,3}$} Julia Kristin Hufenbach,\textit{$^{2,3}$} and Gianaurelio Cuniberti\textit{$^{1,4,*}$}}} \vspace{0.5cm}

 \noindent\small{\textit{$^{1}$~Institute for Materials Science and Max Bergmann Center of Biomaterials, TUD Dresden University of Technology, 01062, Dresden, Germany. }}\\[0.6em]
  \noindent\small{\textit{$^{2}$~Technische Universität Bergakademie Freiberg, Institute for Materials Science, 09599 Freiberg, Germany. }}\\[0.6em]
   \noindent\small{\textit{$^{3}$~Leibniz Institute for Solid State and Materials Research Dresden, 01069 Dresden, Germany. }}\\[0.6em]
 \noindent\small{\textit{$^{4}$~Dresden Center for Computational Materials Science (DCMS), TUD Dresden University of Technology, 01062 Dresden, Germany.}}\\[1em]
  \noindent{$^{\ast}$~Corresponding authors: Leonardo Medrano Sandonas (\texttt{leonardo.medrano@tu-dresden.de}), Gianaurelio Cuniberti   (\texttt{gianaurelio.cuniberti@tu-dresden.de})}
 \end{center}


 
\clearpage 

\clearpage 

\section*{Supplementary Figures}

\begin{figure}[H]
\centering 
\includegraphics[scale=0.4]{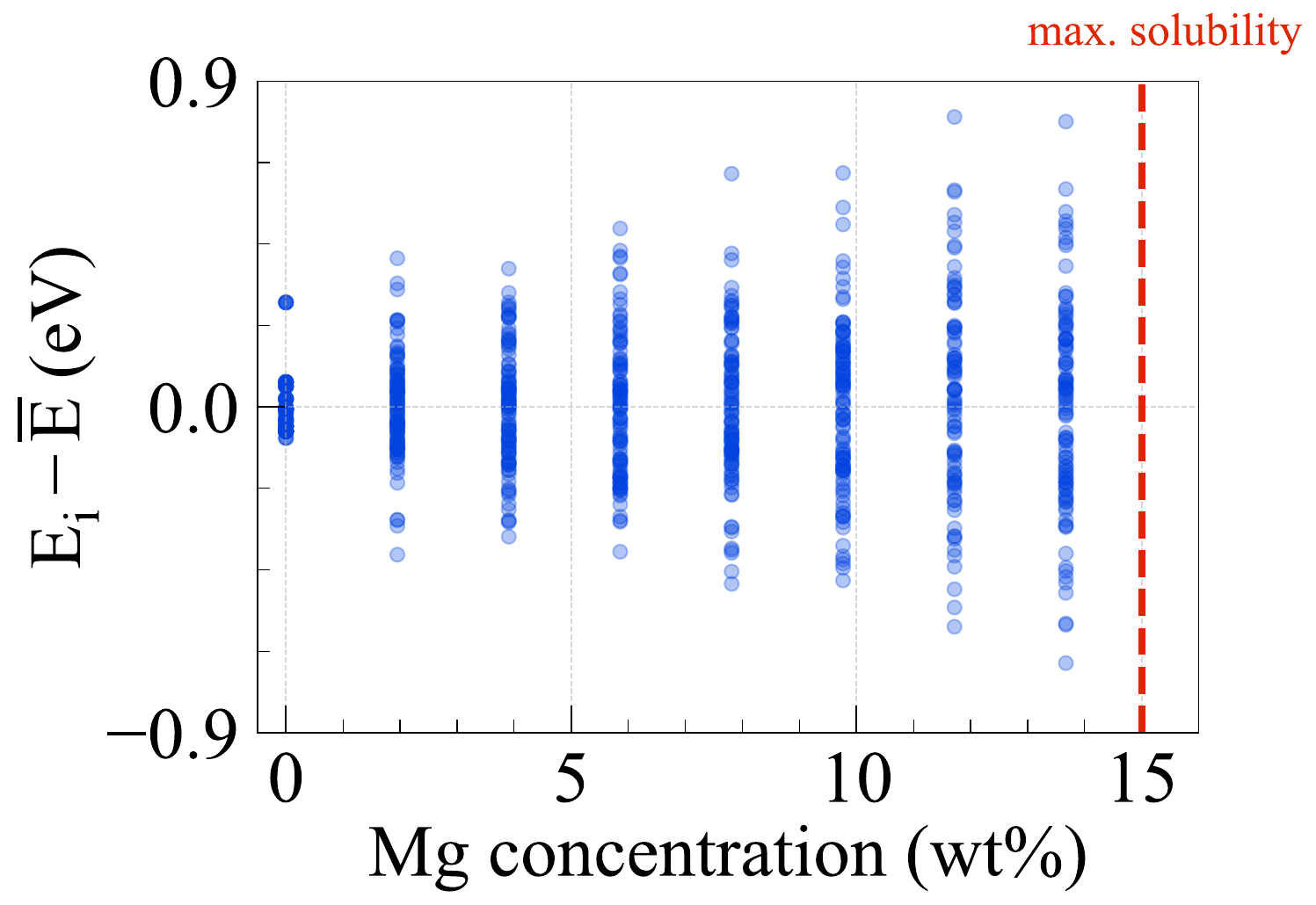}
\caption{\label{fig:SI1} Total energy of preoptimized Al solid solutions (T=0~K) with respect to Mg concentration. For specific concentrations of Mg, 100 different configurations were sampled. The total energy is shifted by the average total energy $\bar{E}$ for the given concentration. 0 wt\% Mg corresponds to the Al alloy with 3 wt\% Zr. Low and high energy configurations are chosen for training.
}
\end{figure}

\begin{figure}[H]
    \centering
    \includegraphics[scale=0.4]{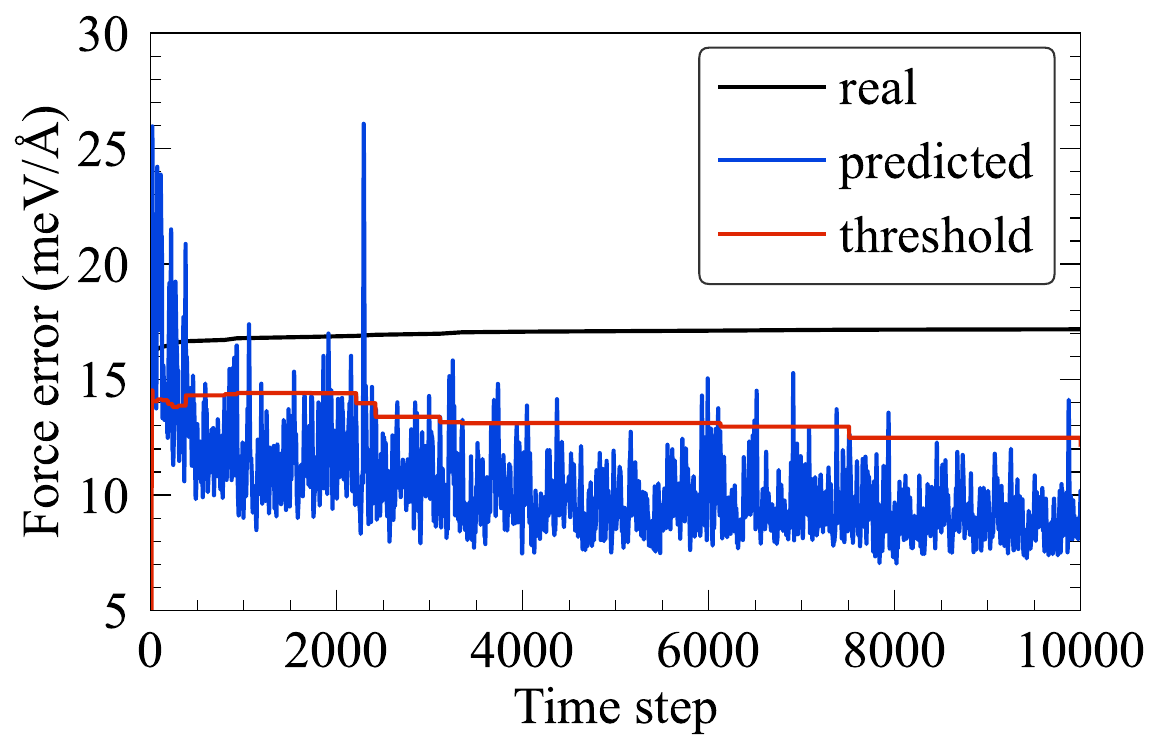}\par
    \includegraphics[scale=0.4]{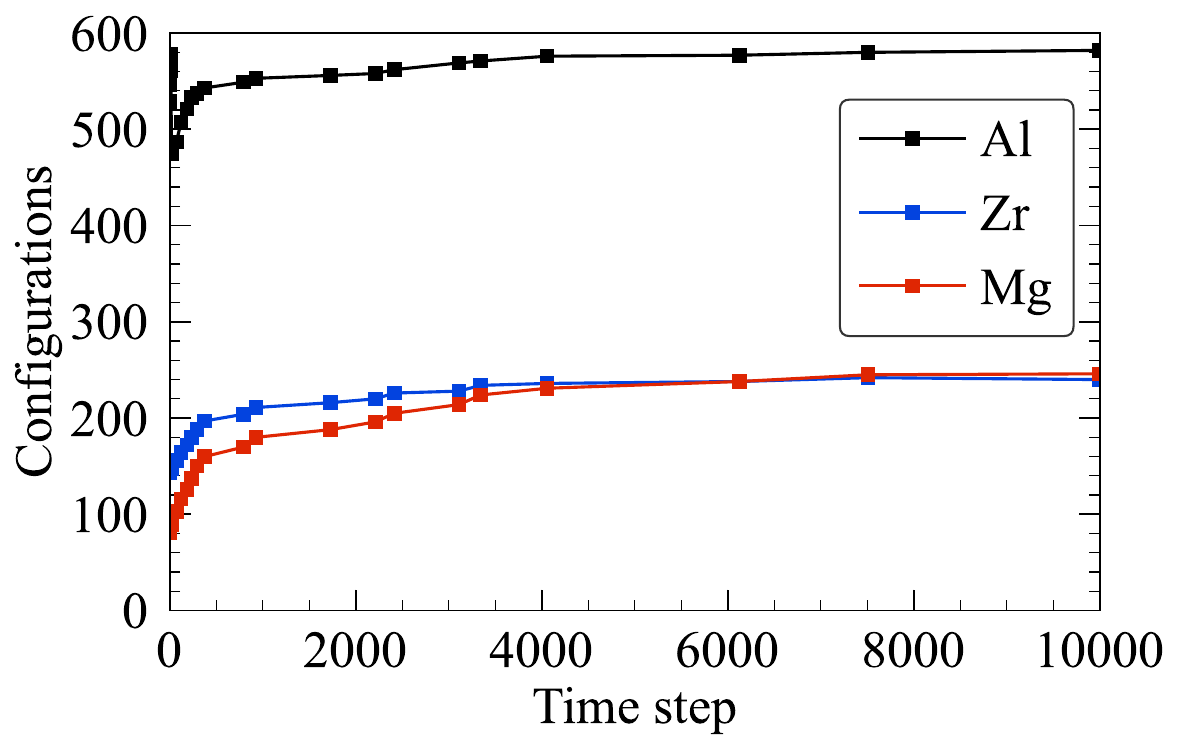}
    \caption{
        Analysis for the training process with VASP-ML of Al$_{249}$Mg$_5$Zr$_2$. 
        Top: force error as a function of time step. The red line shows the Bayesian error threshold, the blue curve the maximum predicted Bayesian error, and the black line the RMSE of predictions vs. ab initio forces. 
        Bottom: Number of local reference configurations for each atom type sampled at each time step.
    }
    \label{fig:training_analysis}
\end{figure}
    \begin{figure}[H]
    \centering 
    \includegraphics[scale=0.3]{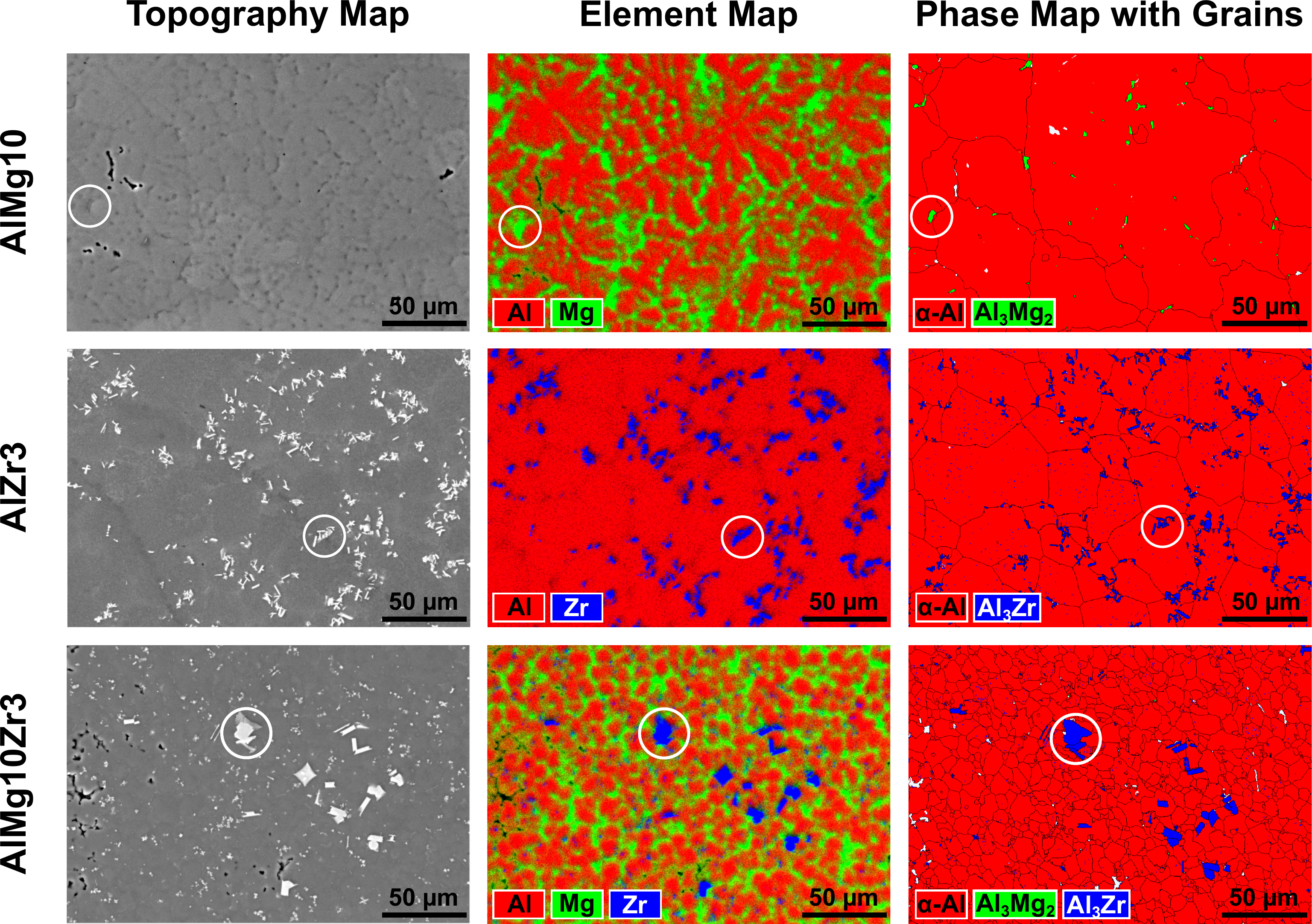}
    \caption{\label{fig:SEM-EDX-EBSD} Topography, elemental and phase map of each alloy system: AlMg10, AlZr3, and AlMg10Zr3. Within the phase maps, the grain boundaries are included in black and white areas represent non-indexed region. The white circles show the same spot on each map for one alloy system. The phase maps are slightly elongated due to the required specimen tilting during measurements. 
    }
    \end{figure}

\newpage

\section*{Supplementary Tables}
   \begin{table}[H]
        \centering
        \caption{Determined density $\rho$ by Archimedes' method and calculated quantities by performing ultrasonic measurements like Poisson's ratio $\nu$, Young's Modulus $E$, Bulk modulus $B$ and shear modulus $G$ depending on the materials composition.} 
        \setlength{\tabcolsep}{4pt}
        \renewcommand{\arraystretch}{1.5}
        \begin{tabular}{cccccc}
        \hline
        Material & $\rho$/~$\frac{g}{cm^3}$ & $\nu$ & $E$/~GPa  & $B$/~GPa  &	$G$/~GPa \\
        \hline
        Al & $2.696\pm{0.001}$ & $0.331\pm{0.007}$ & $72.4\pm{1.4}$ & $71.3\pm{2.2}$ & $27.2\pm{0.6}$ \\ 
        AlMg10 & $2.576\pm{0.010}$ & $0.336\pm{0.008}$ & $66.3\pm{1.7}$ & $67.5\pm{2.3}$ & $24.8\pm{0.8}$ \\ 
        AlZr3 & $2.751\pm{0.007}$ & $0.333\pm{0.001}$ & $73.2\pm{0.5}$ & $73.0\pm{0.8}$ & $27.5\pm{0.2}$ \\ 
        AlMg10Zr3 & $2.599\pm{0.007}$ & $0.326\pm{0.009}$& $67.3\pm{1.6}$ & $64.5\pm{2.4}$ & $25.4\pm{0.7}$ \\ 
        \hline
        \end{tabular}
        \label{tab:Quantities_ultrasonic_measurements}
    \end{table}
\begin{table}[H]
    \centering
    \caption{Local reference configurations $N_B$ within $N_{str}$ snapshots saved as training structures during the simulations in NVT- and NPT-ensemble. The fraction $x$ represents the numbers of first principle calculations, which were skipped during the training. } 
    \setlength{\tabcolsep}{2pt}
    \renewcommand{\arraystretch}{1.5}
    \begin{tabular}{cccccc}
    \hline
    Ensemble & $x$(\%) & $N_{str}$ & $N_B$   \\
    \hline
     NVT&98.2&2693 & 3997 (Al)   \\ 
     && &3179 (Mg)     \\
     & & & 1286 (Zr)   \\
    NPT & 97.8 & 2688	& 3652 (Al)  \\
     &	 &  &2812 (Mg)   \\
     & &  & 1297 (Zr)	  \\
    \hline
    \end{tabular}
    \label{tab:trainingstructures}
\end{table}
\begin{table}[H]
    \centering
    \caption{RMSE for the whole stress tensor $\tau_{ij} $ on the test set with  $\omega_\tau = 10$ in VASP-ML, NVT-ensemble.  } 
    \setlength{\tabcolsep}{2pt}
    \renewcommand{\arraystretch}{1.5}
    \begin{tabular}{ccccccc}
    \hline
    & xx & yy & zz & xy& yz&zx   \\
    \hline
     RMSE (GPa) &0.014&0.012 & 0.014 &0.024&0.021&0.023   \\ 

    \hline
    \end{tabular}
    \label{tab:performance-test-sets}
\end{table}
\begin{table}[H]
    \centering
    \caption{RMSE in energies, forces and stresses evaluated for six different MACE models are shown. The first three models are trained without stresses, whereas the last three models where trained including the stress (indicated by MACE-ML-Xs). The stress tensor contains six independent entries and the RMSE for the largest entry is shown. The units for energy, forces and stresses are meV/atom, eV/\AA \, and GPa, respectively. Next to it, the hyperparameters for the training of each model are shown. The best model is indicated in bold and is used for the calculation of elastic properties.   } 
    \setlength{\tabcolsep}{2pt}
    \renewcommand{\arraystretch}{1.5}
    \begin{tabular}{cccccccccccc}
    \toprule
&& Training data & &&Test data & & & Hyperparameters\\
\midrule
\textrm{Model}&

\textrm{E}&
\textrm{F}&
\textrm{$\tau$}&
\textrm{E}&
\textrm{F}&
\textrm{$\tau$}&
Layers&
Channels&
$L_{max}$&
Corr.order&
$R_{cut}$\\
\midrule
  MACE-ML-1& 6.9& 0.017& 0.21  & 2.7 & 0.026 & 2.39&2&128 &2&3&6 \AA\\
    MACE-ML-2& 5.5&0.017&0.22 & 2.6& 0.026 & 2.31 &2&128&2&3&5 \AA\\
   MACE-ML-3&3.3&0.027&0.30& 1.6 &0.041 & 1.98&2&128 &2&3&4 \AA\\
    \textbf{MACE-ML-4s}&  \textbf{2.3}&\textbf{0.016}&\textbf{0.18} &\textbf{2.7} & \textbf{0.026} & \textbf{0.19}&\textbf{2}&\textbf{96}&\textbf{2}&\textbf{3} & \textbf{6 \AA}\\  
   MACE-ML-5s &6.1 &0.017&  0.22 & 2.9 & 0.026 & 0.28&2 &96&2&3&5 \AA\\

  MACE-ML-6s&  5.4&0.018&0.24 &2.8 & 0.028 & 0.42&2&96&2&3 &4 \AA\\
    \bottomrule
    \end{tabular}
    \label{tab:rmse}
\end{table}
    \begin{table}[H]
    \centering
    \caption{Elastic properties comparison between machine learning models from VASP and MACE and experimental results from ultrasonic measurements. The units of $B$, $G$ and $E$ are GPa.} 
    \setlength{\tabcolsep}{4pt}
    \renewcommand{\arraystretch}{1.5}
    \begin{tabular}{ccccccccccccc}
    \toprule
    & \multicolumn{4}{c}{\textbf{VASP-ML}} & \multicolumn{4}{c}{\textbf{MACE-ML}} & \multicolumn{4}{c}{\textbf{Experiments}} \\
    \midrule
     & Al & AlMg10 & AlZr3 & AlMg10Zr3 & Al & AlMg10 & AlZr3 & AlMg10Zr3 & Al & AlMg10 & AlZr3 & AlMg10Zr3 \\
    \midrule
    $B$      & 72.8 & 66.6 & 68.9 & 65.1 & 76.1 & 69.7 & 76.6 & 71.6 & 71.3 & 67.5 & 73.0 & 64.5 \\
    $G$     & 27.7 & 24.2 & 25.9 & 22.9  & 26.9 & 25.1 & 24.9 & 24.1  & 27.2 & 24.8 & 27.5 & 25.4 \\
    $E$     & 73.7 & 64.9 & 68.9 & 61.2 & 72.2 & 67.2 & 67.5 & 65.0 & 72.4 & 66.3 & 73.2 & 67.3 \\
    $\nu$         & 0.33 & 0.34 & 0.33 & 0.34 & 0.34 & 0.34 & 0.35 & 0.35 & 0.33 & 0.34 & 0.33 & 0.33 \\
    \bottomrule
    \end{tabular}
    \label{tab:elastic_comparison}
\end{table}
\begin{table}[H]
    \centering
    \caption{Chemical formulas of the structures for different supercell sizes by maintaining the weigth percentages for the experimental structures.  } 
    \setlength{\tabcolsep}{2pt}
    \renewcommand{\arraystretch}{1.5}
    \begin{tabular}{cccccc}
    \hline
  wt\%  & $4\times4\times4$ & $5\times5\times5$ & $6\times6\times6$& $7\times7\times7$& $8\times8\times8$   \\
    \hline
    Al& \ch{Al256} &\ch{Al500}&\ch{Al864} &\ch{Al1372} &\ch{Al2048}\\ 
AlMg10 &\ch{Al228Mg28}&\ch{Al445Mg55}&\ch{Al769Mg95}&\ch{Al1221Mg151}&\ch{Al1823Mg225}\\
AlZr3 &\ch{Al254Zr2}&\ch{Al496Zr4}&\ch{Al856Zr7}&\ch{Al1360Zr12}&\ch{Al2032Zr17}\\
AlMg10Zr3 &\ch{Al226Mg28Zr2}&\ch{Al441Mg55Zr4}&\ch{Al763Mg94Zr7}&\ch{Al1206Mg154Zr12}&\ch{Al1802Mg229Zr17}\\
    \hline
    \end{tabular}
\end{table}
\newpage

\section*{Supplementary Derivations}

\subsection*{Training method: Potential energy fitting}
%

The method is implemented in the Vienna \textit{Ab initio} Simulation Package (VASP) and described by Jinnouchi \textit{et al.}. For a deeper insight the reader is referred to their work since only the most important ideas are recalled here. The main assumption of the approach is to approximate the total potential energy $U$ of $N_A$ atoms  in the system as a sum over local atomic energies $U_i$
\begin{eqnarray}
  U = \sum_{i=1}^{N_A}U_i.
\label{eq:one}
\end{eqnarray}
Atomic energies are given as a linear combination of kernels $K$ as 
\begin{eqnarray}
    U_i = \sum_{i_B=1}^{N_B} w_{i_B} K(\mathbf{x}_i, \mathbf{x}_{i_B}),
\label{eq:two}    
\end{eqnarray}
where summation runs over $N_B$ reference atoms, which are chosen on the fly. The kernel measures the similarity of two atoms through their corresponding descriptors $\mathbf{x}_i$ and $\mathbf{x}_{i_B}$. Two types of descriptors are used distinguished by two- and three-body distribution functions. The two-body distribution function $\rho_i^{(2)}$ is defined as the probability to find an atom $j \neq i $ at distance $r$ from atom $i$ as 
\begin{eqnarray}
    \rho_i^{(2)}(r) = \frac{1}{4\pi} \int \dif \mathbf{\hat{r}} \,  \rho_i(r \mathbf{\hat{r}}).
\label{eq:three}
\end{eqnarray}
Here, $\mathbf{\hat{r}}$ is the unit vector along the distance $r$ between two atoms and $\int \dif \mathbf{\hat{r}} = \int \dif \phi \int \dif \theta \sin \theta$ integrates over the angular part of a sphere, while $r$ stays constant. The smoothed three-dimensional distribution function of the i-th atom is given by 
\begin{eqnarray}
 \rho_i (\mathbf{r}) &&= \sum_{j \neq i}^{N_R} \tilde{\rho}_{ij}(\mathbf{r}), \\
  \tilde{\rho}_{ij}(\mathbf{r})  &&= f_{\mathrm{cut}}(|\mathbf{r}_i - \mathbf{r}_j|)  g(\mathbf{r}- (\mathbf{r}_i-\mathbf{r}_j)), \label{eq:five}
\end{eqnarray}
where the cutoff function $f_{\mathrm{cut}}$ takes account of $N_R$ atoms within a sphere with radius $R_{\mathrm{cut}}$ and $g(\mathbf{r})$ is a smoothed $\delta$-function. The three-body distribution function $\rho_i^{(3)}$ is defined as the probability to find an atom $j \neq i $ at distance $r$ from atom $i$ and another atom $k \neq i,j$ at distance $s$ from atom $i$ and the spanning angle $\vartheta = \angle (\mathbf{r},\mathbf{s}) $ 
\begin{eqnarray}
    \rho_i^{(3)}(r,s,\vartheta) = &&\int \dif \mathbf{\hat{r}}  \int \dif \mathbf{\hat{s}} \, \delta(\mathbf{\hat{r}}\cdot \mathbf{\hat{s}} - \cos \vartheta) \nonumber \\
    &&\times \sum_{j \neq i }^{N_R}  \sum_{k \neq i,j}^{N_R}  \tilde{\rho}_{ij}(r\mathbf{\hat{r}})  \tilde{\rho}_{ik}(s\mathbf{\hat{s}}).
\label{eq:six}
\end{eqnarray}
Both, the two- and three-body distribution function in the form of Eq.\ \ref{eq:three} and \ref{eq:six} obey translational and rotational invariance. In practise, they are discretized by expanding them into orthonormalized spherical Bessel functions $\chi_{nl}$ and Legendre polynomials $P_l$ up to certain orders $N_R^0$, $N_R^l$ and $L_{\mathrm{max}}$ as
\begin{eqnarray}
    \rho_i^{(2)}(r) &&= \frac{1}{\sqrt{4\pi}} \sum_{n=1}^{N_R^0} c_n^i \chi_{n0}(r), \\
    \rho_i^{(3)}(r,s,\vartheta) &&= \sum_{l=0}^{L_{\mathrm{max}}} \sum_{n=1}^{N_R^l} \sum_{\nu=1}^{N_R^l} \sqrt{\frac{2l+1}{2}} p_{n\nu l}^i \chi_{nl}(r) \chi_{\nu l}(s) \nonumber \\
    && \quad  \times P_l(\cos \vartheta). 
\end{eqnarray}
Eventually, the descriptor of the $i$-th atom $\mathbf{x}_i=(\mathbf{x}_i^{(2)},\mathbf{x}_i^{(3)})$ is given by the coefficients 
\begin{eqnarray}
    \mathbf{x}_i^{(2)} &&= \sqrt{\beta}(c_1^i, \dots,c_{N_R^0}^i)^{\trans} , \\
    \mathbf{x}_i^{(3)} &&= \sqrt{1 - \beta}(p_{110}^i , \dots, p_{N_R^{L_{\mathrm{max}}}N_R^{L_{\mathrm{max}}}L_{\mathrm{max}}}^i)^{\trans}.
\end{eqnarray}
The hyperparameter $\beta$ is controlling the weight of the two- and three-body descriptors. In this model, a polynomial Kernel of the form
\begin{eqnarray}
    K(\mathbf{x}_i,\mathbf{x}_j ) = ( \mathbf{\hat{x}}_i \cdot \mathbf{\hat{x}}_j)^\zeta
\end{eqnarray}
is used. The exponent $\zeta$ introduces mixing of two- and three-body descriptors and it has been shown that mixing leads to a more accurate MLIP, especially for liquid systems. \\ 
Machine learning atomic forces and the stress tensor are calculated from derivatives of the potential energy given by Eq.\ \ref{eq:two}. Hence, the potential energy, atomic forces, and stress tensor for the $\alpha$-th time step of an AIMD simulation are aligned to a $1+ 3 N_A + 6$ dimensional vector $\mathbf{y}^{\alpha} = (E^{\alpha}, \{\mathbf{F}_i^\alpha\}, \boldsymbol{\tau}^\alpha )  $ and Eq. \ref{eq:one} and \ref{eq:two} can be written as a matrix equation
\begin{eqnarray}
    \mathbf{y}^\alpha = \boldsymbol{\phi}^\alpha \cdot \mathbf{w},
\label{eq:twelve}    
\end{eqnarray}
where $\boldsymbol{\phi}^\alpha$ is a $(1+3N_A+6) \times N_B$ matrix containing terms like $ \sum_i K(\mathbf{x}_i^\alpha, \mathbf{x}_{i_B})$ and the corresponding derivatives. By introducing the super vector $\mathbf{Y} = (\mathbf{y}^1, \dots, \mathbf{y}^{N_{\mathrm{str}}})$ the linear problem Eq.\ \ref{eq:twelve} is generalized to 
\begin{eqnarray}
    \mathbf{Y} = \boldsymbol{\Phi} \cdot \mathbf{w},
\end{eqnarray}
where $\boldsymbol{\Phi}$ is called design matrix. The $N_B$ linear coefficients are aligned to the vector $\mathbf{w}$. The desired optimal coefficients are calculated by Bayesian linear regression from all $N_{\mathrm{str}}$ structures  $\{\mathbf{y}^\alpha\} $ by maximizing the posterior conditional probability distribution
\begin{eqnarray}
    p(\mathbf{w}|\mathbf{Y}) = \frac{p(\mathbf{Y}|\mathbf{w}) p(\mathbf{w})}{p(\mathbf{Y})}.
\end{eqnarray}
The conjugate prior probability is assumed to be a zero mean Gaussian with covariance $\sigma_w^2 \mathbf{I}$, i.e.\ $p(\mathbf{w}) = \mathcal{N}(\mathbf{w}| \mathbf{0}, \sigma_w^2 \mathbf{I})$. Secondly, the likelihood function is assumed to be a Gaussian with covariance $\sigma_v^2 \mathbf{I}$, i.e.\ $p(\mathbf{Y}|\mathbf{w}) = \mathcal{N}(\mathbf{Y}| \boldsymbol{\Phi} \cdot \mathbf{w},\sigma_v^2 \mathbf{I}) $. The normalization of the posterior distribution is governed by $p(\mathbf{Y})$. It follows 
\begin{align}
    p(\mathbf{w}|\mathbf{Y}) &= \mathcal{N}(\mathbf{w} | \mathbf{m}, \mathbf{S}), \\
    \mathbf{m} &= \frac{1}{\sigma_v^2} \mathbf{S} \cdot \boldsymbol{\Phi}^{\trans} \cdot \mathbf{Y}, \\
    \mathbf{S}^{-1} &= \frac{1}{\sigma_w^2}\mathbf{I} + \frac{1}{\sigma_v^2} \boldsymbol{\Phi}^{\trans} \cdot  \boldsymbol{\Phi}.
\end{align}
The posterior is maximized at the center of the Gaussian distribution, thus the solution for the coefficients is given by $\mathbf{m}$ calculated via LU-factorization. The covariance parameters $\sigma_v$ and $\sigma_w$ are optimized by evidence approximation .\\
\\
\textbf{Hyperparameters of VASP-ML:}
For the radial and angular descriptors, a radial cutoff $R_{\mathrm{cut}} = 5 $~\AA \, and spreading of $0.5 $~\AA \,for $g(\mathbf{r})$, see Eq.\ \ref{eq:five}, is used. The number of radial basis functions was set to $N^0_R = N^l_R = 6$ and the maximum angular momentum quantum number is $L_{\mathrm{max}}=6$. The mixing parameter for two- and three-body descriptors was set to $\zeta = 4$ and $\beta = 0.5$. \\
\subsection*{Error calculation of polycrystalline constants}
The moduli are functions of stiffness constants, \ie $M = M(\{C_{ij}\})$ (replace $M$ by the corresponding modulus). We calculate the error by propagation of uncertainty as 
\begin{equation*}
    u(M) = \sqrt{\sum_{i,j} \left( \frac{\partial M}{\partial C_{ij}}\right)^2 u^2(C_{ij})},
\end{equation*}
where the uncertainty of $C_{ij}$ is given by
\begin{equation*}
    u(C_{ij}) = \frac{1}{N} \sqrt{\sum_{k,l} u^2(c_{kl})}.
\end{equation*}
$N = 3$ for $C_{11}$ and $C_{44}$ and $N = 6$ for $C_{12}$, also received by error propagation of Eq.\ 5 in the main text. The constants $c_{kl}$ are received from OLS of the stress vs strain curve for the different stretch states as $\tau_k = c_{kl} \varepsilon_l$. For simplicity, we suppress the indices for the stiffness matrix and corresponding directions of stress and strain, $\tau= c \varepsilon$. We apply strains up to $\pm1\%$ as described in the main text, and each of them is now labeled as $\varepsilon^{(m)}$. The total number of strain states is either five or nine (including $\varepsilon=0$). The corresponding stress of a strain rate is given by $\tau^{(m)}$ and we calculate it as an average from the last 500 steps of the corresponding MLIP MD simulations. The error of $c$ is then given by the standard error of the slope as 
\begin{equation*}
   u(c) =  SE(c) = \sqrt{\frac{\hat\sigma^2}{\sum_m (\varepsilon^{(m)}- \bar{\varepsilon})^2}},
\end{equation*}
where $m$ runs over the total number of strain states $n$ (five or nine) and $\bar{\varepsilon}= 0 $, because positive and negative strain rates cancel out. Furthermore, $\hat{\sigma}^2$ is the residual variance given by 
\begin{equation*}
    \hat{\sigma}^2 = \frac{1}{n-2} \sum_m r_m^2 , 
\end{equation*}
where $r_m = \tau^{(m)} -  (c\varepsilon^{(m)}+ \tau(\varepsilon=0))$ are the residuals for the stress. 
\newpage
